\documentclass{article}
\usepackage{graphics}
\usepackage{bm}
\usepackage{graphicx}
\usepackage{amsmath}
\usepackage{amssymb}
\usepackage{amstext}
\usepackage{amscd}
\usepackage{amsfonts}
\usepackage{hyperref}
\usepackage{textcomp}
\usepackage{color}
\DeclareMathAlphabet{\EuFrak}{U}{euf}{m}{n}
\DeclareMathAlphabet{\EuScript}{U}{eus}{m}{n}

\newcommand{\be}{\begin{equation}}
\newcommand{\ee}{\end{equation}}
\newcommand{\ben}{\begin{eqnarray}}
\newcommand{\een}{\end{eqnarray}}

\newcommand{\nd}{\noindent}

\date{\today}

\begin{document}

\title{{\bf Gravitational partition function modified by super-light brane world perturbative modes}}
\author{{M. Hameeda$^{1,2,a}$, B. Pourhassan$^{3,b}$, M.C.Rocca$^{4,5,6,c,d}$}\\
{\texttt{\rm{ Aram Bahroz Brzo$^{7,e}$ }}}\\
\small{$^1$ Department of Physics, Government Degree College, Tangmarg, Kashmir, 193402 India}\\
\small{$^2$ Inter University Centre for Astronomy and Astrophysics , Pune India}\\
\small{$^3$ School of Physics, Damghan University,}\\
\small{ P. O. Box 3671641167, Damghan, Iran}\\
\small{$^4$ Departamento de F\'{\i}sica,
Universidad Nacional de La Plata,}\\
\small{$^5$ Departamento de Matem\'{a}tica,
Universidad Nacional de La Plata,}\\
\small{$^6$ Consejo Nacional de Investigaciones Cient\'{\i}ficas
y Tecnol\'{o}gicas}\\
\small{(IFLP-CCT-CONICET)-C. C. 727, 1900 La Plata -
Argentina}\\
\small{$^7$ Department of Physics, College of Education,}\\
\small{ University of Sulaimani, Kurdistan-Iraq}\\
\small{\texttt{\rm{$^{a}$hme123eda@gmail.com, $^{b}$b.pourhassan@du.ac.ir,}}}\\
\small{\texttt{\rm{$^{c}$rocca@fisica.unlp.edu.ar,$^{d}$mariocarlosrocca@gmail.com, }}}\\
\small{\texttt{\rm{$^{e}$aram.brzo@univsul.edu.iq}}}}

\date{}

\maketitle

\begin{abstract}
\nd
In this paper, we will analyze the effects of super-light brane world perturbative modes on clustering of galaxies.
In present manuscript, we use the Boltzmann and Tsallis statistical approaches to study the large distance modification of gravity in the brane world. The impact of modified potential on clustering of galaxies is analyzed in both the approaches.
The infinities associated with Newtonian point mass approximation of gravity models is removed using the analytical extensions.
The regularized and finite partition function is obtained for the system of galaxies in the brane world model and is used to evaluate the regularized thermodynamic properties of the system in the form of equations of state. Hence, we study thermodynamic quantities and discuss about the thermodynamic stability of the model. We find that large number of galaxies may lead to the thermodynamic instability.

\nd The appealing feature of the paper revolve around the generalization of the dimensional regularization (GDR) of Bollini and Giambiagi supplemented with the statistical adequacy of Boltzmann and Tsallis.

\end{abstract}

\newpage

\tableofcontents

\newpage

\renewcommand{\theequation}{\arabic{section}.\arabic{equation}}

\section{Introduction}

\nd
In this paper, we will analyze a large distance  modification to the gravitational partition function. This gravitational partition function has been used in analyzing the large scale structure of our universe, and hence any large modification of the gravitational potential would modify it.

\nd
Clustering of galaxies and the structure formation in the expanding universe using Newtonian potential as a source of attraction and by approximating the galaxies as point masses has been extensively analyzed \cite{sas}; \cite{sas84}; \cite{sas2}. The study has been followed and improved to analyze the structure formation for extended objects \cite{ahm06}; \cite{ahm10}; .

\nd
An empirical modification of Newtonian dynamics(MOND), has been proposed and discussed at low accelerations  \cite{mil,mil1,mil2}.
Gravitational clustering with modified potentials have been studied in various modifications of Newtonian potential to explain clustering \cite{Hameeda}; \cite{hameeda16}; \cite{Ham20}; \cite{4}; \cite{1}; \cite{1b};
The modified theories like the non-local extensions of GR can be a possible treatment for the singularities like black hole, big bang and the cosmic expansion \cite{capo,capo1,capo2,capo3,modes}.

\nd In most of the studies the standard techniques of statistical mechanics have been followed \cite{sut}; \cite{sut1}. In our early studies we have analyzed the clustering of galaxies using a Newtonian potential modified by superlight modes of a brane world model and obtained considerable effects on the clustering parameter and have elaborated the effects on cosmic energy equations \cite{Hameeda}. The large distance corrections to the Newtonian potential from super-light modes in brane world models have been studied by considering the universe a brane in higher dimensions \cite{bro}. The string theory due to the inherent characteristics of possessing extra dimensions thus, become the motivation to discuss these models \cite{ran}. Due to the lack of testing of general relativity and the corresponding Newtonian approximation at very large or very small distances and due to the propagation of gravity in to higher dimensional bulk, the Newtonian potential gets brane corrections \cite{ark}; \cite{ark1}; \cite{ant}. Keeping in consideration the importance of models with super-light perturbation modes and the modification of gravitational interaction at astronomical scales, the form of Newtonian potential used in this paper to study the clustering can be a promising candidate to resolve the issue of dark matter in galaxies \cite{bro}.

\nd
It may be noted that most of the work done on clustering of galaxies, using this formalism, has been done using the Boltzmann-Gibbs entropy.  However, it can be argued that as  the   Boltzmann-Gibbs entropy  is based on the extensive property of the system,  it might not explain the gravitational systems, as these systems can violate the extensive property  \cite{t1}. Such a  violation occurs because of   breakdown of ergodicity for such gravitational systems  \cite{t2,t4}. Thus, it has been suggested that such system can be better described by  Tsallis statistical mechanics \cite{t, t8}. In fact, it has been demonstrated that the    Tsallis statistical mechanics can be used to study  self interacting gravitational particles, even though such systems can violate the extensive property  \cite{t6}. So, we will also analyze the Tsallis statistical mechanics for this system of galaxies corrected by  super-light brane world perturbative modes.

\nd
Though in our early studies we have analyzed the brane world modified Newtonian potential to galactic clustering using the techniques of statistical mechanics and have further modified the potential using the softening parameter to eliminate the mathematical divergences. In present study we use the Boltzmann statistical as well as Tsallis statistical approaches to study the gravitating gas interacting through brane potential. Dimensional regularization of partition function to get the divergence free results is the main thrust involved in the paper. The paper involves rigorous work on dimensional regularization to regularize the thermodynamic properties in both the approaches. Besides giving physical impact of the large distance modification of gravity in the brane world, the mathematics of the paper is quite appealing due to the generalization of the dimensional regularization of Bollini and Giambiagi \cite{xz12, xz14, prd}. This generalization was based on the general quantification method of QFT's \cite{la12, la14, la16, la18} using Ultradistributions of Sebastiao e Silva, also known as Ultrahyperfunctions \cite{jss, hasumi, tp8}.

Our analysis is valid if the following assumptions are satisfied: \\

\nd
1) \color{red}The graviton is massless,\normalcolor as suggested by the LIGO experiment.
In that case the graviton does not present polarizations that produce
strong corrections to the garvitational potential, given for example by
positive powers of r as shown in reference \cite{dva}.
This was proved in the reference \cite{jpco1,cam}, in which gravity is quantized following the paths suggested by Suraj N. Gupta and Richard P. Feynman, through the theory of ultrahyperfunctions \cite{la12,la14,la16,la18}. This theory allows to quantize Non-Renormalizable Quantum Field Theories. \\

\nd
2) The \color{red} graviton is non-massless.\normalcolor In that case the result of the reference \cite{dva} shows that there is a polarization of the graviton that produces a strong correction that is a positive power of r. So, for our result to be valid, the sources must be beyond the radius below which the additional polarizations of the graviton are strongly interacting.
\normalcolor

\section{Partition function}

Let's consider the distribution
$\frac {1} {r}=PV\frac {1} {r}$.
Then $\frac {1} {r}\mid_{r=0}=0$.
The general partition function of a system of $N$ particles of mass $m$ interacting through the modified gravitational potential
with potential energy is $\Phi$,  can be written as
\begin{eqnarray}
\label{eq2.1}
{\cal Z}_\nu=\frac{1}{N!}\int d^{\nu}pd^{\nu}r
\times  \exp\biggl(-\biggl[\sum_{i=1}^{N}\frac{p_{i}^2}{2m}+\Phi(r_{1}, r_{2}, r_{3},
\dots, r_{N})\biggr] T^{-1}\biggr),
\end{eqnarray}
The interaction potential energy between galaxies in brane world model with large extra dimensions,
can be written as \cite{ran}; \cite{cal}

\begin{equation}
\label{eq2.2}
\phi_{i,j}=-\frac{Gm^2}{ r_{ij}}\left(1+\frac{k_m}{ r_{ij}^2}\right),
\end{equation}
For $N$ galaxies the potential energy can be expressed as
\begin{equation}
\label{eq2.3}
\phi_{i,j}=-\frac{N(N-1)Gm^2}{ 2r_{ij}}\left(1+\frac{k_m}{ r_{ij}^2}\right),
\end{equation}
The partition function ${\cal Z}$ in $\nu$ dimensions.
\begin{equation}
\label{eq2.4}
{\cal Z}_\nu=\frac{1}{N!}\int\limits_{-\infty}^{\infty}d^\nu x\int\limits_{-\infty}^{\infty}d^\nu p
\exp\left[\beta\left(\frac {N(N-1)Gm^2} {2r}\left(1+\frac{k_m}{ r^2}\right)-\frac {Np^2} {2m}\right)\right]
\end{equation}
Evaluating the integral over the angles we have:
\[{\cal Z}_\nu=\frac{1}{N!}\left(\frac {2\pi^{\frac {\nu} {2}}} {\Gamma\left(\frac {\nu} {2}\right)}\right)^2
\int\limits_0^{\infty}r^{\nu-1}dr\exp\left(\beta\frac {N(N-1)Gm^2} {2r}\left(1+\frac{k_m}{r^2}\right)\right)\times\]
\begin{equation}
\label{eq2.5}
\int\limits_0^{\infty}p^{\nu-1} dp\exp\left(-\frac {Np^2} {2m}\right)
\end{equation}
For the integral over the momentum we obtain:
\begin{equation}
\label{eq2.6}
\int\limits_0^{\infty}p^{\nu-1}dr\exp\left(-\beta\left(\frac {Np^2} {2m}\right)\right)=
\frac {\Gamma\left(\frac {\nu} {2}\right)(2m)^{\frac {\nu} {2}}} {2(\beta N)^{\frac {\nu} {2}}}
\end{equation}
To solve the integral over $r$
\[\int\limits_0^{\infty}r^{\nu-1}dr\exp\left(\beta\frac {N(N-1)Gm^2} {2r}\left(1+\frac{k_m}{r^2}\right)\right)\]
we make use of the approximation:
\[\int\limits_0^{\infty}x^{-\nu-1}dx\exp\left(\beta_1x+\beta_2{x^3}\right)=\int\limits_0^{\infty}x^{-\nu-1}dx\exp\left(\beta_1x\right)\left(1+\beta_2{x^3}\right)\]
Thus we get:
\[\int\limits_0^{\infty}x^{-\nu-1}dx\exp\left(\beta_1x\right)+\beta_2\int\limits_0^{\infty}x^{-\nu+2}dx\left(\exp\left(\beta_1x\right)\right)=\beta_1^\nu\left[\frac{1}{-\nu(1-\nu)(2-\nu)}+\frac{\beta_2}{\beta_1^3}\right]\Gamma\left(3-\nu\right)\]
Using this result, we have for ${\cal Z}$ the expression:

\begin{eqnarray}\label{eq2.7}
{\cal Z}_\nu&=&\frac{1}{N!}\frac {2\pi^\nu }{\Gamma\left(\frac {\nu} {2}\right)}
\left(\frac{2m} {\beta N}\right)^{\frac {\nu} {2}}
\left(\beta\frac {N(N-1)Gm^2} {2}\right)^\nu\nonumber\\
&\times&\left[\frac{1}{-\nu(1-\nu)(2-\nu)}+k_m\left(\beta\frac {N(N-1)Gm^2} {2}\right)^{-2}\right]\Gamma\left(3-\nu\right)
\end{eqnarray}

\begin{figure}[h!]
 \begin{center}$
 \begin{array}{cccc}
\includegraphics[width=40 mm]{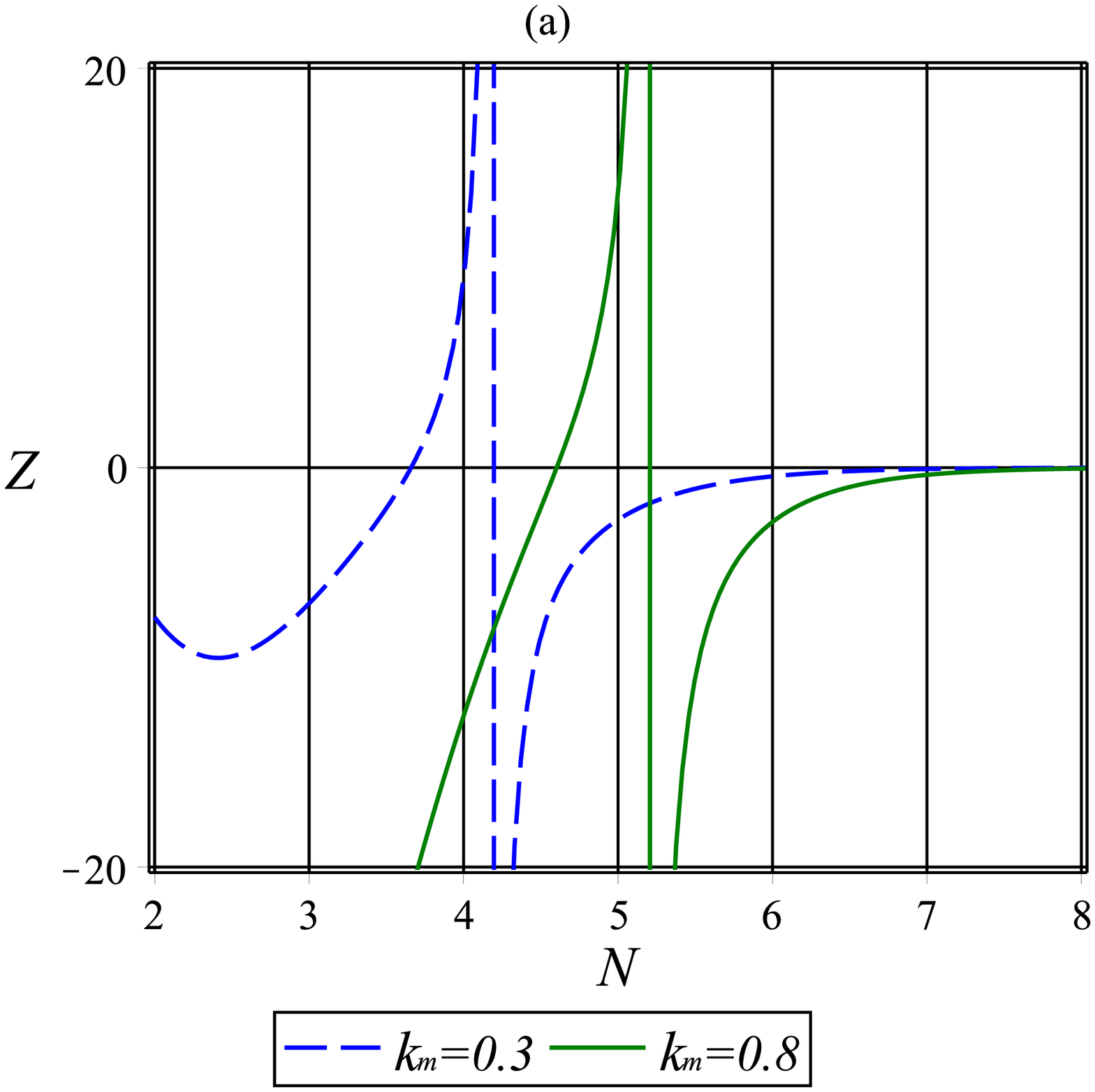}
\includegraphics[width=40 mm]{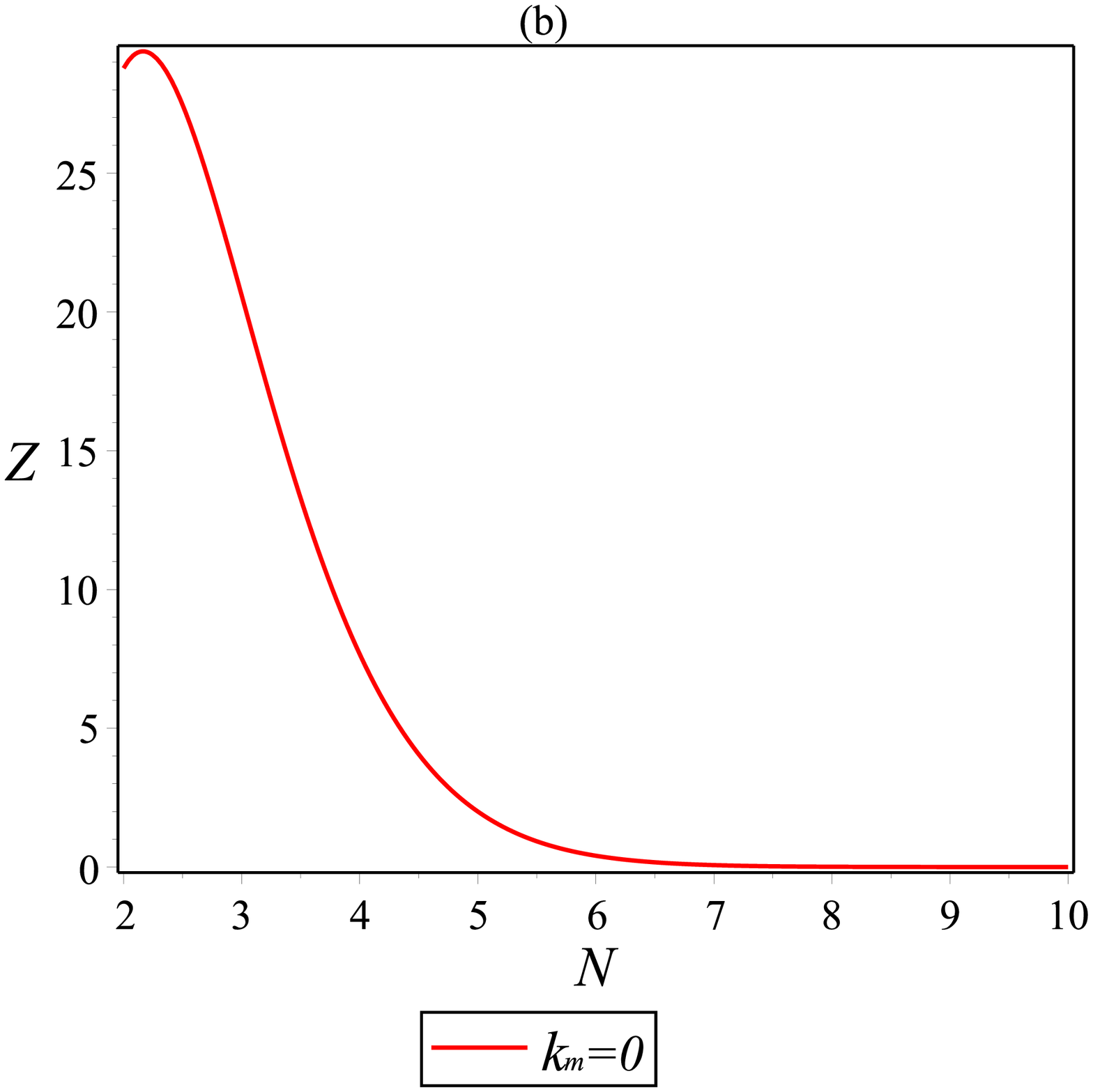}
\includegraphics[width=40 mm]{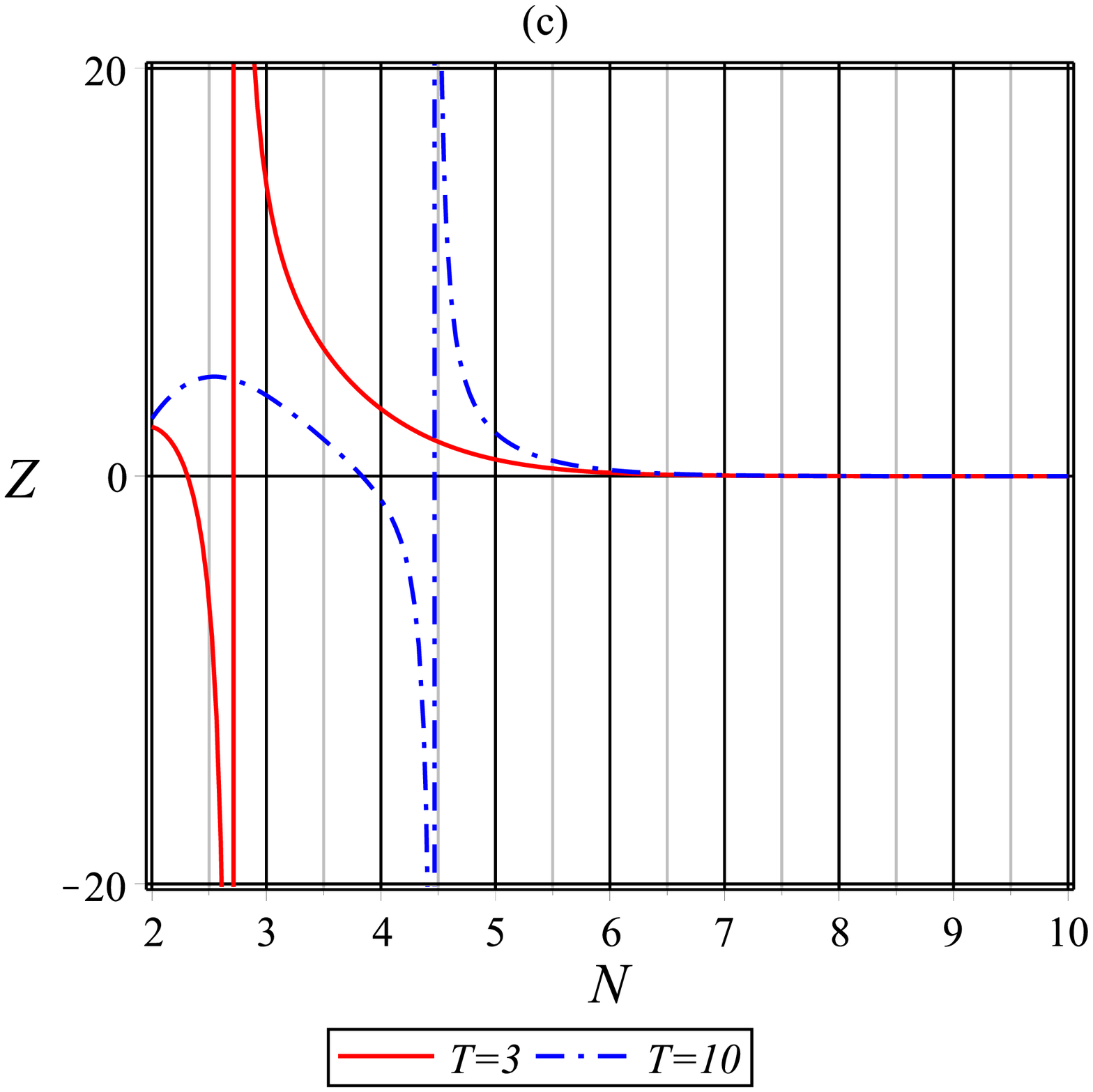}
 \end{array}$
 \end{center}
\caption{Partition function in the brane world model with $G=m=1$. (a) $T=5$; (b) $T=3$; (c) $k_{m}=0.1$.}
 \label{fig1}
\end{figure}

\nd Now, we proceed to do Laurent series development around $ \nu = 3 $. We write:
\begin{equation}
\label{eq2.8}
{\cal Z}_\nu=f(\nu)\Gamma(3-\nu)
\end{equation}
where
\[f(\nu)=\frac{1}{N!}\frac {2\pi^\nu }{\Gamma\left(\frac {\nu} {2}\right)}
\left(\frac{2m} {\beta N}\right)^\frac {\nu} {2}
\left(\beta\frac{N(N-1)Gm^2}{2}\right)^\nu\times\]
\begin{equation}
\label{eq2.9}
\left[\frac{1}{-\nu(1-\nu)(2-\nu)}+k_m\left(\beta\frac {N(N-1)Gm^2}{2}\right)^{-2}\right]
\end{equation}
In three dimensions it is:
\begin{equation}
\label{eq2.10}
f(3)=\frac{1}{N!}{4\pi^{\frac{5}{2}}}
\left(\frac{2m} {\beta N}\right)^\frac {3} {2}\left[-\frac{1}{6}\alpha+k_m\alpha\right],
\end{equation}
where:
\begin{equation}
\label{eq2.11}
\alpha=\beta\frac{N(N-1)Gm^2}{2}
\end{equation}
We get:
\[\frac{f^{'}(\nu)}{f(\nu)}=\frac{\nu}{2}\psi\left(\frac{\nu}{2}\right)+\frac{1}{2}\ln\left(\frac{\pi^2\beta N(N-1)^2G^2m^5}{2}\right)+\]
\begin{equation}
\label{eq2.12}
\frac{-2+6\nu-3\nu^2}{-\nu(1-\nu)(2-\nu)+k_m\alpha^{-2}\nu^2(1-\nu)^2(2-\nu)^2}
\end{equation}
In three dimensions (\ref{eq2.12}) takes the form:
\begin{equation}
\label{eq2.13}
\frac{f^{'}(3)}{f(3)}=\frac{3}{2}\psi\left(\frac{3}{2}\right)+\frac{1}{2}\ln\left(\frac{\pi^2\beta N(N-1)^2G^2m^5}{2}\right)+\frac{11}{6}\left(\frac{\alpha^2}{\alpha^2-6k_m}\right)
\end{equation}
For the $\Gamma$ function we have
\begin{equation}
\label{eq2.14}
\Gamma(3-\nu)=\frac {1} {3-\nu}-C+\sum\limits_{k=1}^\infty c_k\left(\nu-3\right)^k
\end{equation}
As a consequence the partition function in $\nu$ dimensions is:
\begin{equation}
\label{eq2.15}
{\cal Z}_\nu=\frac {f(3)} {3-\nu}-f(3)C-f^{'}(3)+\sum\limits_{k=1}^{\infty}a_k\left(\nu-3\right)^k
\end{equation}
Thus:
\begin{equation}
\label{eq2.16}
{\cal Z}=-f(3)\left(C+\frac{f^{'}(3)}{f(3)}\right)
\end{equation}
For three dimensions it is explicitly expressed as:
\[{\cal Z}=-\frac{4}{N!}{\pi^{\frac{5}{2}}}
\left(\frac{2m} {\beta N}\right)^\frac {3}{2}\left[-\frac{1}{6}\alpha+k_m\alpha\right]\]
\begin{equation}
\label{eq2.17}
\left(C+\frac{3}{2}\psi\left(\frac{3}{2}\right)+\frac{1}{2}\ln\left(\frac{\pi^2\beta N(N-1)^2G^2m^5}{2}\right)+\frac{11}{6}\left(\frac{\alpha^2}{\alpha^2-6k_m}\right)\right)
\end{equation}
Finally after simplification we arrive at the expression of finite partition function:
\begin{eqnarray}\label{eq2.18}
{\cal Z}&=&\left(6-C+\ln\left(\frac{\pi^2\beta N(N-1)^2G^2m^5}{128}\right)+\frac{11}{3}\left(\frac{\alpha^2}{\alpha^2-6k_m}\right)\right)\nonumber\\
&\times&\frac{2\alpha}{N!}{\pi^{\frac{5}{2}}}
\left(\frac{2m} {\beta N}\right)^\frac {3}{2}\left(\frac{1}{6}-k_m\right).
\end{eqnarray}
In Fig. \ref{fig1} we can see behavior of the partition function for various values of the model parameters. The most important point here is to have positive partition function which is corresponding to the physical system. It help us to constrain model parameters. We can use this fact to restrict values of $N$ and $k_{m}$. First of all we should note that interacting system need $N\geq2$. Hence, in the case of $k_m>\frac{1}{6}$, partition function is mostly negative except for special number of galaxies like $N=4$ or $N=5$ (see Fig. \ref{fig1} (a)). On the other hand, for the case of $k_m<\frac{1}{6}$ we can obtain a lower bound for the number of galaxies. In the special case of $k_{m}=0$ the partition function is completely positive for $N\geq2$ (see Fig. \ref{fig1} (b) where drawn for $T=3$ however other temperatures yield to the similar result). Finally, in Fig. \ref{fig1} (c) we can see $0<k_m<\frac{1}{6}$ yields to positive partition function if $N>N_{min}$. Value of $N_{min}$ is depend on the temperature. Increasing temperature, increased value of $N_{min}$.

\section{Thermodynamic properties from Boltzmann statistics}

\setcounter{equation}{0}

\nd
We know from Boltzmann Gibbs statistical mechanics and thermodynamics, that there are several physical quantities like average energy $U$, Helmholtz free energy $F$, entropy $S$ and chemical potential $\mu$, that can be of interest to a physical system. Therefore, we dedicate this section to demonstrate how we can calculate various thermodynamic properties of the gravitational system in question. These quantities may not have the same meaning as in usual thermodynamics since these are for a gravitational system, nonetheless, their physical significance is intact. As we shall see, these quantities will be crucial in further sections.

\nd
We start with calculating the mean energy $<{\cal U}>$ of a system at constant $N,V$ by using the formula:
\begin{equation}
\label{eq3.1}
<{\cal U}>=-\frac {1} {{\cal Z}}\frac {\partial{\cal Z}} {\partial\beta}
\end{equation}
Then, we get for $<{\cal U}>$:
\begin{equation}
\label{eq3.2}
<{\cal U}>=\frac{1}{\beta}-\frac{1}{\beta}\left(\frac{1-44\left(\frac{k_m\alpha^2}{(\alpha^2-6k_m)^2}\right)}{-{C}-6\ln 2+6+\ln\left(\frac{\pi^2\beta N(N-1)^2G^2m^5}{2}\right)+\frac{11}{3}\left(\frac{\alpha^2}{\alpha^2-6k_m}\right)}\right).
\end{equation}


\nd It help us to obtain the specific heat via
\begin{equation}
\label{eq3.3.3}
C_v=\left(\frac{\partial U}{\partial T}\right)_V,
\end{equation}
which yields to the following expression,
\[C_v=k+\frac{k\left(\frac{176\alpha^2k_m}{(\alpha^2-6k_m)^3}-\frac{44\alpha^2k_m}{(\alpha^2-6k_m)^2}-1\right)}{-{C}-6\ln 2+6+\ln\left(\frac{\pi^2\beta N(N-1)^2G^2m^5}{2}\right)+\frac{11}{3}\left(\frac{\alpha^2}{\alpha^2-6k_m}\right)}\]
\begin{equation}
\label{eq3.4}
-\frac{k\left(1-\left(\frac{44k_m\alpha^2}{(\alpha^2-6k_m)^2}\right)^2\right)}{\left(-{C}-6\ln 2+6+\ln\left(\frac{\pi^2\beta N(N-1)^2G^2m^5}{2}\right)+\frac{11}{3}\left(\frac{\alpha^2}{\alpha^2-6k_m}\right)\right)^2}
\end{equation}
Thus
we see that the system is self-gravitating, as established by Lynden Bell in \cite{lynden}

\nd Helmholtz free energy $F$ is obtained using
\begin{equation}
\label{eq3.5}
F=-\frac{1}{\beta}\ln {\cal Z}
\end{equation}
Therefore we get:
\[F=-\frac{1}{\beta}\biggl[\ln\pi+N-N\ln N+\ln(\alpha-6k_m\alpha)+\]
\begin{equation}
\label{eq3.6}
\ln\left(18-3C-18\ln 2+3\ln\left(\frac{\pi^2\beta N(N-1)^2G^2m^5}{2}\right)+\frac{11\alpha^2}{\alpha^2-6k_m}\right)\biggr]
\end{equation}


The entropy $S$ is obtained from
\begin{equation}
\label{eq3.7}
TS=U-F
\end{equation}
Its explicit expression is:
\[S=k-k\left(\frac{1-44\left(\frac{k_m\alpha^2}{(\alpha^2-6k_m)^2}\right)}{-{C}-6\ln 2+6+\ln\left(\frac{\pi^2\beta N(N-1)^2G^2m^5}{2}\right)+\frac{11}{3}\left(\frac{\alpha^2}{\alpha^2-6k_m}\right)}\right)\]\[+k\biggl[\ln\pi+N-N\ln N+\ln(\alpha-6k_m\alpha)+\]
\begin{equation}
\label{eq3.8}
\ln\left(18-3C-18\ln 2+3\ln\left(\frac{\pi^2\beta N(N-1)^2G^2m^5}{2}\right)+\frac{11\alpha^2}{\alpha^2-6k_m}\right)\biggr]
\end{equation}
Analyzing the entropy shows positive region for $N_{min}\leq N\leq N_{max}$ in the case of $0<k_m<\frac{1}{6}$, hence we find upper limit for the number of galaxies which is strongly depend on the temperature.

\begin{figure}[h!]
 \begin{center}$
 \begin{array}{cccc}
\includegraphics[width=100mm]{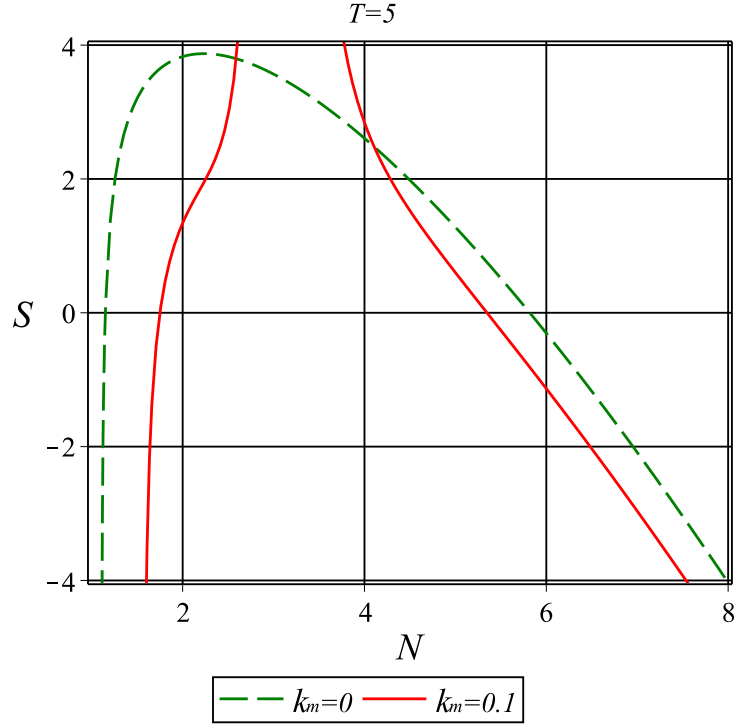}
 \end{array}$
 \end{center}
\caption{Entropy of Boltzmann statistics with $G=m=1$, $k_m=1/6$.}
 \label{fig5}
\end{figure}

The chemical potential is defined as:
\begin{equation}
\label{eq3.9}
\mu=\left(\frac{\partial F}{\partial N}\right)_T
\end{equation}
Accordingly:
\begin{equation}
\label{eq3.10}
\mu=-\frac{1}{\beta}\left[-\ln N+\frac{2N-1}{N(N-1)}+\frac{\frac{3(3N-1)}{N(N-1)}+\frac{2N-1}{N(N-1)}\left(\frac{22\alpha^2}{\alpha^2-6k_m}-\frac{22\alpha^4}{(\alpha^2-6k_m)^2}\right)}{-{C}-6\ln 2+6+\ln\left(\frac{\pi^2\beta N(N-1)^2G^2m^5}{2}\right)+\frac{11}{3}\left(\frac{\alpha^2}{\alpha^2-6k_m}\right)}\right]
\end{equation}

\begin{figure}[h!]
 \begin{center}$
 \begin{array}{cccc}
\includegraphics[width=100 mm]{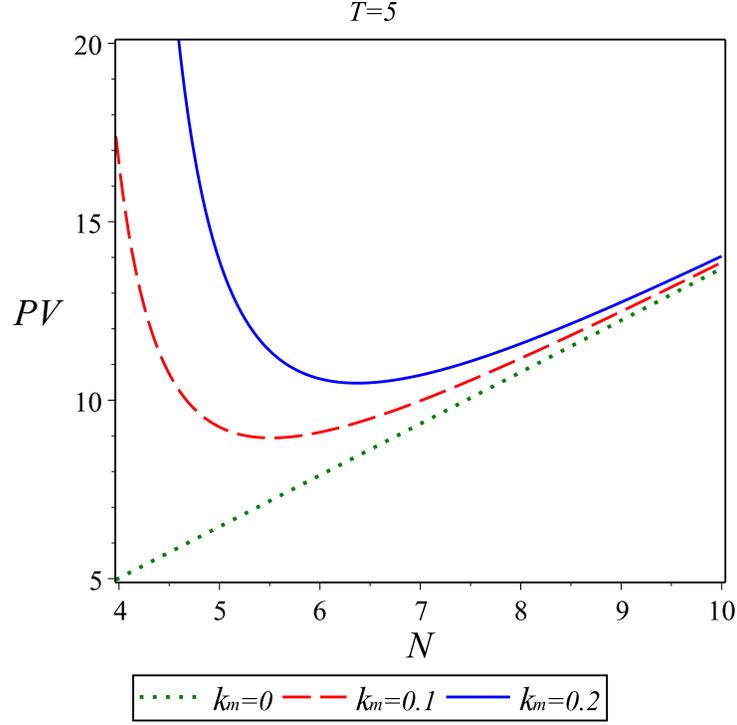}
 \end{array}$
 \end{center}
\caption{EoS of Boltzmann statistics with $G=m=1$.}
 \label{fig7}
\end{figure}

\nd In addition, we have the following formula, which we will use later:
\begin{equation}
\label{eq3.11}
\exp{\beta N\mu}=N^N\exp{\left[\frac{2N-1}{(N-1)}+\frac{\frac{3(3N-1)}{(N-1)}+\frac{2N-1}{(N-1)}\left(\frac{22\alpha^2}{\alpha^2-6k_m}-\frac{22\alpha^4}{(\alpha^2-6k_m)^2}\right)}{-{C}-6\ln 2+6+\ln\left(\frac{\pi^2\beta N(N-1)^2G^2m^5}{2}\right)+\frac{11}{3}\left(\frac{\alpha^2}{\alpha^2-6k_m}\right)}\right]}
\end{equation}
Pressure $P$ is obtained using
\begin{equation}
\label{eq3.12}
PV=\frac{2N}{3}<{\cal U}>
\end{equation}
Using (\ref{eq3.2}) we arrive to the state equation:
\begin{equation}
\label{eq3.13}
\beta PV=\frac{2N}{3}-\frac{2N}{3}\left(\frac{1-44\left(\frac{k_m\alpha^2}{(\alpha^2-6k_m)^2}\right)}{-{C}-6\ln 2+6+\ln\left(\frac{\pi^2\beta N(N-1)^2G^2m^5}{2}\right)+\frac{11}{3}\left(\frac{\alpha^2}{\alpha^2-6k_m}\right)}\right)
\end{equation}
In Fig. \ref{fig7} we can see equation of state (EoS) for various interaction strength. In the case of $k_{m}=0$ we can see linear behavior. For the large $N$ limit, there is no important interaction effect.

\subsection{Probability distribution function}

\nd Up to this point, we have dealt with the microcanonical ensemble. However, now we shall define the grand canonical partition function $Z_G$ to calculate the probability distribution of the system. We define $Z_G$ as follows

\begin{equation}
\label{eq4.1}
\ln Z_G=\beta PV
\end{equation}

\nd From chemical potential $\mu$ we find the fugacity $z$ as
\begin{equation}
\label{eq4.2}
z^N=e^{N\beta\mu}
\end{equation}
The distribution function follows as:
\begin{equation}
\label{eq4.3}
F(N)=\frac{z^N{\cal Z}}{Z_G}
\end{equation}
Using (\ref{eq3.11}), the distribution function can be written as:
\[F(N)=\frac{2N^N{\pi^{\frac{5}{2}}}}{N!}\left(\frac{2m} {\beta N}\right)^\frac {3}{2}\left[\frac{1}{6}\alpha-k_m\alpha\right]\]
\[\exp\biggl[\frac{2N-1}{(N-1)}-\frac{2N}{3}+\frac{\frac{3(3N-1)}{(N-1)}+\frac{2N-1}{(N-1)}\left(\frac{22\alpha^2}{\alpha^2-6k_m}-\frac{22\alpha^4}{(\alpha^2-6k_m)^2}\right)+\frac{2N}{3}\left(1-44\left(\frac{k_m\alpha^2}{(\alpha^2-6k_m)^2}\right)\right)}
{-C-6\ln 2+6+\ln\left(\frac{\pi^2\beta N(N-1)^2G^2m^5}{2}\right)+\frac{11}{3}\left(\frac{\alpha^2}{\alpha^2-6k_m}\right)}\]
\begin{equation}
\label{eq4.4}
\left(-{C}-6\ln 2+6+\ln\left(\frac{\pi^2\beta N(N-1)^2G^2m^5}{2}\right)+\frac{11}{3}\left(\frac{\alpha^2}{\alpha^2-6k_m}\right)\right)
\end{equation}


\section{Tsallis partition function of the system in the brane world model}

\setcounter{equation}{0}

\nd
In 1988, C. Tsallis proposed a non-extensive generalization of statistical mechanics (widely called as Tsallis statistical mechanics) \cite{t1}. Based on the motivations laid out in the introduction, we now try to investigate the many-body gravitational system using Tsallis's formalism.
Tsallis q-exponential is defined as the distribution:
\begin{equation}
\label{eq5.1}
e_q(x)=[1+(q-1)x]_+^{\frac {1} {q-1}}
\end{equation}
Or equivalently
\begin{equation}
\label{ep5.2}
e_q(x)=
\begin{cases}
1+(q-1)x]^{\frac {1} {q-1}}\;\;\;;\;\;\;1+(q-1)x>0\\
0\;\;\;;\;\;\;1+(q-1)x<0
\end{cases}
\end{equation}
Let's consider again the distribution:
$\frac {1} {r}=PV\frac {1} {r}$.
First we do the calculation in $\nu$ dimensions and assume  $q>1$. Thus
\begin{equation}
\label{eq5.3}
{\cal Z}_\nu=\int\limits_{-\infty}^{\infty}d^\nu x\int\limits_{-\infty}^{\infty}d^\nu p
\left[1+(q-1)\beta\left(\frac {N(N-1)Gm^2} {2r}+\frac{kN(N-1)Gm^2}{2r^3}-\frac {Np^2} {2m}\right)\right]_+^{\frac {1} {q-1}}
\end{equation}
Here we put $q=4/3$. The choice of that value of $q$ is not arbitrary. It is the value of $q$ where Verlinde's conjecture of emergent gravity could be proved, in the non-relativistic case \cite{di14}. Accordingly:
\begin{equation}
\label{eq5.4}
{\cal Z}_\nu=\left[\frac {2\pi^{\frac {\nu} {2}}} {\Gamma\left(\frac {\nu} {2}\right)}\right]^2
\int\limits_0^{\infty}r^{\nu-1}dr\int\limits_0^{\infty}p^{\nu-1} dp
\left[1+\beta\left(\frac {N(N-1)Gm^2} {6r}+\frac{N(N-1)Gm^2}{6r^3}-\frac {Np^2} {6m}\right)\right]_+^{3}
\end{equation}
The expression for ${\cal Z}$ in (\ref{eq5.4}) can be rewritten in the form:
\begin{equation}
\label{eq5.5}
{\cal Z}_\nu=\left[\frac {2\pi^{\frac {\nu} {2}}} {\Gamma\left(\frac {\nu} {2}\right)}\right]^2\left(\frac{6m}{\beta N}\right)^{\frac{\nu}{2}}B(4,\frac{\nu}{2})\int\limits_{0}^{\infty}r^{-\frac{\nu}{2}-10}dr\left(r^3+\alpha_t r+k\alpha_t\right)^{\frac{\nu}{2}+3}
\end{equation}
where
\[\alpha_t=\frac{\beta N(N-1)Gm^2}{6}\]
Hence we can write
\begin{equation}
\label{eq5.6}
{\cal Z}_\nu=\left[\frac {2\pi^{\frac {\nu} {2}}} {\Gamma\left(\frac {\nu} {2}\right)}\right]^2\left(\frac{6m}{\beta N}\right)^{\frac{\nu}{2}}B(4,\frac{\nu}{2})\int\limits_{0}^{\infty}r^{-\frac{\nu}{2}-10}dr(r+b)^{\frac{\nu}{2}+3}(r-r_1)^{\frac{\nu}{2}+3}(r-r_3)^{\frac{\nu}{2}+3}
\end{equation}
where $r_1$, $r_2=-b$, and $r_3$  are the cube roots of the integrand.
Also,
\[r_1=\gamma_1+e^{\frac{i\pi}{3}}\gamma_2\]
\[r_3=\gamma_1+e^{-\frac{i\pi}{3}}\gamma_2\]
\[b=\gamma_2-\gamma_1-\frac{\alpha}{3}\]
\[\gamma_1=\left(\sqrt{\frac{1}{4}\left(\frac{2\alpha_t^3}{27}+\alpha_t k\right)^2+\frac{1}{27}\left(\frac{\alpha}{3}(\alpha_t-6)\right)^2}-\frac{1}{2}\left(\frac{2\alpha_t^3}{27}+\alpha_t k\right)\right)^{\frac{1}{3}}\]
\[\gamma_2=\left(\sqrt{\frac{1}{4}\left(\frac{2\alpha_t^3}{27}+\alpha_t k\right)^2+\frac{1}{27}\left(\frac{\alpha}{3}(\alpha_t-6)\right)^2}+\frac{1}{2}\left(\frac{2\alpha_t^3}{27}+\alpha_t k\right)\right)^{\frac{1}{3}}\]
After solving the integral we have

\begin{eqnarray}\label{eq5.7}
{\cal Z}_\nu&=&\left[\frac {2\pi^{\frac {\nu} {2}}} {\Gamma\left(\frac {\nu} {2}\right)}\right]^2\left(\frac{6m}{\beta N}\right)^{\frac{\nu}{2}}B(4,\frac{\nu}{2})b^{\nu}\frac{\Gamma\left(-\frac{\nu}{2}-9\right)}{\Gamma\left(-\frac{3\nu}{2}-9\right)}\nonumber\\
&\times&\Gamma(-\nu){}F\left(-\nu,-\frac{\nu}{2}-3,-\frac{\nu}{2}-3,-\frac{3\nu}{2}-9;1+\frac{r_1}{b};1+\frac{r_3}{b}\right).
\end{eqnarray}
Here, $F$ is the hypergeometric function of two variables.
After a long calculation we get (See Appendix  A):
\[{\cal Z}_\nu=\frac {24\pi^\nu} {\Gamma\left(\frac {\nu} {2}\right)^2}\left(\frac{6m}{\beta N}\right)^{\frac{\nu}{2}}b^{\nu}\frac{\Gamma\left(-\frac{\nu}{2}-9\right)}{(3+\frac{\nu}{2})(2+\frac{\nu}{2})(1+\frac{\nu}{2})\Gamma\left(-\frac{3\nu}{2}-9\right)}\]
\begin{equation}
\label{5.15}
\left(\Gamma(3-\nu)h_\nu+\phi_\nu(z_1,z_2)\right)
\end{equation}

\subsection{Evaluation of partition function}

Now we do the Laurent development explicitly, taking into account that $\phi_\nu(z_1,z_2)$ does not have a pole at $\nu = 3$ and take the expression for $\phi_3$ at the end of the development.
Thus the Laurent development go as:\\
We can write
\begin{equation}
\label{eq6.1}
{\cal Z}_\nu=\Gamma(3-\nu)f_\nu+g_\nu\phi_\nu(z_1,z_2)
\end{equation}
where
\begin{equation}
\label{eq6.2}
f_\nu=\frac {24\pi^\nu} {\Gamma\left(\frac {\nu} {2}\right)^2}\left(\frac{6m}{\beta N}\right)^{\frac{\nu}{2}}b^{\nu}\frac{\Gamma\left(-\frac{\nu}{2}-9\right)}{(3+\frac{\nu}{2})(2+\frac{\nu}{2})(1+\frac{\nu}{2})\Gamma\left(-\frac{3\nu}{2}-9\right)}h_\nu
\end{equation}
\begin{equation}
\label{eq6.3}
f_\nu=\frac {24\pi^\nu} {\Gamma\left(\frac {\nu} {2}\right)^2}\left(\frac{6m}{\beta N}\right)^{\frac{\nu}{2}}b^{\nu}\frac{\Gamma\left(\frac{3\nu}{2}+10\right)\sin{\frac{3\nu\pi}{2}}}{(3+\frac{\nu}{2})(2+\frac{\nu}{2})(1+\frac{\nu}{2})\Gamma\left(\frac{\nu}{2}+10\right)\sin{\frac{\pi\nu}{2}}}h_\nu
\end{equation}

\begin{equation}
\label{eq6.4}
g_\nu=\frac {24\pi^\nu} {\Gamma\left(\frac {\nu} {2}\right)^2}\left(\frac{6m}{\beta N}\right)^{\frac{\nu}{2}}b^{\nu}\frac{\Gamma\left(\frac{3\nu}{2}+10\right)\sin{\frac{3\nu\pi}{2}}}{(3+\frac{\nu}{2})(2+\frac{\nu}{2})(1+\frac{\nu}{2})\Gamma\left(\frac{\nu}{2}+10\right)\sin{\frac{\pi\nu}{2}}}
\end{equation}
\begin{equation}
\label{eq6.5}
f_\nu=f_3+f^{'}_3(\nu-3)+\sum\limits_{k=2}^\infty b_k(\nu-3)^k
\end{equation}
For the $\Gamma$ function we have
\begin{equation}
\label{eq6.6}
\Gamma(3-\nu)=\frac {1} {3-\nu}-C+\sum\limits_{k=1}^\infty c_k\left(\nu-3\right)^k
\end{equation}As a consequence:
\begin{equation}
\label{eq6.7}
{\cal Z}_\nu=\frac {f_3} {3-\nu}-f_3C-f^{'}_3+\sum\limits_{k=1}^{\infty}a_k\left(\nu-3\right)^k+g(3)\phi_3(s)
\end{equation}
Now we write partition function as:
\begin{equation}
\label{eq6.8}
{\cal Z}=-f_3\left(C+\frac{f^{'}_3}{f_3}\right)+g_3\phi_3(z_1,z_2)
\end{equation}
Where:
\begin{equation}
\label{eq6.9}
f_3=-\frac{256\pi^2}{105}\left(\frac{6m}{\beta N}\right)^{\frac{3}{2}}b^3\frac{\Gamma(\frac{29}{2})}{\Gamma(\frac{23}{2})}h_3
\end{equation}
And
\begin{equation}
\label{eq6.10}
g_3=-\frac{256\pi^2}{105}\left(\frac{6m}{\beta N}\right)^{\frac{3}{2}}b^3\frac{\Gamma(\frac{29}{2})}{\Gamma(\frac{23}{2})}
\end{equation}
We get:
\begin{equation}
\label{eq6.11}
\frac{f^{'}_3}{f_3}=\frac{1}{2}\ln{\frac{6m\pi^2b^2}{\beta N}}-2\psi(\frac{3}{2})+\psi(\frac{29}{2})-\psi(\frac{23}{2})-\frac{143}{315}+\frac{h^{'}_3}{h_3}
\end{equation}
And also:
\begin{equation}
\label{eq6.12}
h_3=\left(-\frac{1}{6}+\frac{1}{6}(z_1+z_2)-\frac{1}{75}(18z_1z_2+7z_1^2+7z_2^2)+\frac{21}{31050}(5z_1^3+5z_2^3-27z_1^2z_2-27z_1z_2^2)\right)
\end{equation}
\begin{equation}
\label{eq6.13}
h_3^{'}=\left(\frac{11}{36}-\frac{1}{4}(z_1+z_2)+\frac{454}{1875}(z_1z_2)+\frac{171}{1875}(z_1^2+z_2^2)+\frac{62}{198375}(z_1^3+z_2^3)+\frac{3088}{991875}(z_1^2z_2+z_1z_2^2)\right)
\end{equation}
Finally we obtain the finite expression for partition function which is divergence free and write it as:
\begin{equation}
\label{eq6.14}
{\cal Z}=-g_3\biggl[h_3\left(C+\frac{f^{'}_3}{f_3}\right)-\phi_3(z_1,z_2)\biggr]
\end{equation}

\subsection{Mean Energy in $\nu$ dimensions}

We now proceed to evaluate the average energy of the system. For it is:
\[{\cal Z}<{\cal U}>_\nu=\left[\frac {2\pi^{\frac {\nu} {2}}} {\Gamma\left(\frac {\nu} {2}\right)}\right]^2
\int\limits_0^{\infty}r^{\nu-1}dr
\int\limits_0^{P_0} p^{\nu-1} dp
\left(-\frac {N(N-1)Gm^2} {2r}-\frac {kN(N-1)Gm^2} {2r^2}+\frac {Np^2} {2m}\right)\]
\begin{equation}
\label{eq7.1}
\left[1+(q-1)\beta\left(\frac {N(N-1)Gm^2} {2r}+\frac{kN(N-1)Gm^2}{2r^3}-\frac{Np^2}{2m}\right)\right]^{\frac {1} {q-1}}
\end{equation}
Or, equivalently:
\[{\cal Z}<U>_\nu=\biggl[\frac{2\pi^\frac{\nu}{2}}{\Gamma{\frac{\nu}{2}}}\biggr]^2[\frac{\beta N(q-1)}{2m}]^{\frac{1}{q-1}}\]
\[\left(\int\limits_0^{\infty}r^{\nu-1}dr
\int\limits_0^{P_0}\frac{Np^{\nu+1}}{2m}\left[\frac{1}{\beta N(q-1)}+\frac{(N-1)Gm^2}{2r}+\frac{k(N-1)Gm^2}{2r^3}-\frac{p^2}{2m}\right]^{\frac{1}{q-1}}dp\right.\]
\[\left.-\frac {N(N-1)Gm^2} {2}\int\limits_0^{\infty}r^{\nu-2}dr\int\limits_0^{P_0}p^{\nu-1}\left[\frac{1}{\beta N(q-1)}+\frac{(N-1)Gm^2}{2r}+\frac{k(N-1)Gm^2}{2r^3}-\frac{p^2}{2m}\right]^{\frac{1}{q-1}}dp\right)\]
\begin{equation}
\label{eq7.2}
\left.-\frac {kN(N-1)Gm^2} {2}\int\limits_0^{\infty}r^{\nu-3}dr\int\limits_0^{P_0}p^{\nu-1}\left[\frac{1}{\beta N(q-1)}+\frac{(N-1)Gm^2}{2r}+\frac{k(N-1)Gm^2}{2r^3}-\frac{p^2}{2m}\right]^{\frac{1}{q-1}}dp\right)
\end{equation}
For $q=4/3$ we obtain:
\begin{equation}
\label{eq7.3}
{\cal Z}<U>_\nu=\frac {3} {2\beta}\left[\frac {2\pi^{\frac {\nu} {2}}} {\Gamma\left(\frac {\nu} {2}\right)}\right]^2
\left(\frac {6m} {N\beta}\right)^{\frac {\nu} {2}}
\biggl[-\alpha B(4,\frac{\nu}{2})I_1-k\alpha B(4,\frac{\nu}{2})I_2+B(4,\frac{\nu}{2}+1)I_3\biggr]
\end{equation}
Again,after a long calculation we obtain (See Appendix A):
\begin{equation}
\label{eq7.26}
{\cal Z}<U>_\nu=G_\nu[\Gamma(3-\nu)F_\nu+\Phi_\nu(z_1,z_2)\biggr]
\end{equation}
Where we have defined:
\[\Phi_\nu(z_1,z_2)=\phi_1+\phi_2+\phi_3\]
With:
\begin{equation}
\label{eq7.27}
G_\nu=\frac{3}{2\beta}\frac {24\pi^\nu} {\Gamma\left(\frac {\nu} {2}\right)^2}\left(\frac{6m}{\beta N}\right)^{\frac{\nu}{2}}b^{\nu}\frac{\Gamma\left(\frac{3\nu}{2}+11\right)\sin{\frac{3\nu\pi}{2}}}{(3+\frac{\nu}{2})(2+\frac{\nu}{2})(1+\frac{\nu}{2})\Gamma\left(\frac{\nu}{2}+11\right)\sin{\frac{\pi\nu}{2}}}
\end{equation}
And:
\begin{equation}
\label{eq7.28}
F_\nu=\left[-\alpha h_1(\nu)-k\alpha h_2(\nu)\frac{(\frac{3\nu}{2}+12)(\frac{3\nu}{2}+11)}{(\frac{\nu}{2}+12)(\frac{\nu}{2}+11)}+h_3(\nu)\frac{(\frac{3\nu}{2}+12)(\frac{3\nu}{2}+11)}{(\frac{\nu}{2}+12)(\frac{\nu}{2}+11)}\right]
\end{equation}

\subsection{Calculation of mean energy}

We follow the Laurent development as:
\begin{equation}
\label{eq8.1}
G_\nu F_\nu=G_3F_3+(G_3F_3)^{'}(\nu-3)+\sum\limits_{k=2}^\infty b_k(\nu-3)^k
\end{equation}
For the $\Gamma$ function we have
\begin{equation}
\label{eq8.2}
\Gamma(3-\nu)=\frac {1} {3-\nu}-C+\sum\limits_{k=1}^\infty c_k\left(\nu-3\right)^k
\end{equation}As a consequence:
\begin{equation}
\label{eq8.3}
{\cal Z}<U>_\nu=\frac {G_3F_3} {3-\nu}-G_3F_3C-(G_3F_3)^{'}+\sum\limits_{k=1}^{\infty}a_k\left(\nu-3\right)^k+G_3\Phi(3,s)
\end{equation}
and
\begin{equation}
\label{eq8.4}
{\cal Z}<U>=-G_3F_3\left(C+\frac{(G_3F_3)^{'}}{G_3F_3}\right)+G_3\Phi(3,z_1,z_2)
\end{equation}
Equivalently:
\begin{equation}
\label{eq8.5}
{\cal Z}<U>=-G_3F_3\left(C+\frac{G_3^{'}}{G_3}+\frac{F_3^{'}}{F_3}\right)+G_3\Phi(3,z_1,z_2)
\end{equation}
We evaluate $F_3,F'_3,G_3$ and $G'_3$ in three dimensions:. The result is:
\begin{equation}
\label{eq8.6}
G_3=-\frac{128\pi^2}{35\beta}\left(\frac{6m}{\beta N}\right)^{\frac{3}{2}}b^{3}\frac{\Gamma\left(\frac{31}{2}\right)}{\Gamma\left(\frac{25}{2}\right)}
\end{equation}
\begin{equation}
\label{eq8.7}
F_3=\left[-\alpha h_1(3)-\frac{1023}{675}k\alpha h_2(3)+\frac{1023}{675}h_3(3)\right]
\end{equation}
\begin{equation}
\label{eq8.8}
\frac{G_3^{'}}{G_3}=\frac{1}{2}\ln\left(\frac{6m}{\beta N}\pi^2b^2\right)+\psi(\frac{31}{2})-\psi(\frac{25}{2})-2\psi(\frac{3}{2})-\frac{143}{315}
\end{equation}
\begin{equation}
\label{eq8.9}
F_3^{'}=\left[-\alpha h_1^{'}(3)-\frac{1023}{675}k\alpha h_2^{'}(3)+\frac{1023}{675}h_3^{'}(3)+\frac{3332}{151875}k\alpha h_2(3)-\frac{3332}{151875}h_3(3)\right]
\end{equation}
We get then:
\begin{equation}
\label{eq8.10}
<U>=-\frac {G_3F_3} {{\cal Z}}  \left(C+\frac{G_3^{'}}{G_3}+\frac{F_3^{'}}{F_3}\right)+G_3\Phi(3,z_1,z_2)
\end{equation}
We can see behavior of above energy in terms of temperature and number. 
we can see that $<U>$ is increasing function of $N$. However, 
we shows that behavior of system is different for small and large $N$. For the large $N$ limit we can see energy is decreased by temperature which may be sign of an instability. In order to study thermodynamics stability we need to investigate about the specific heat.\\
\begin{figure}[h!]
 \begin{center}$
 \begin{array}{cccc}
\includegraphics[width=65 mm]{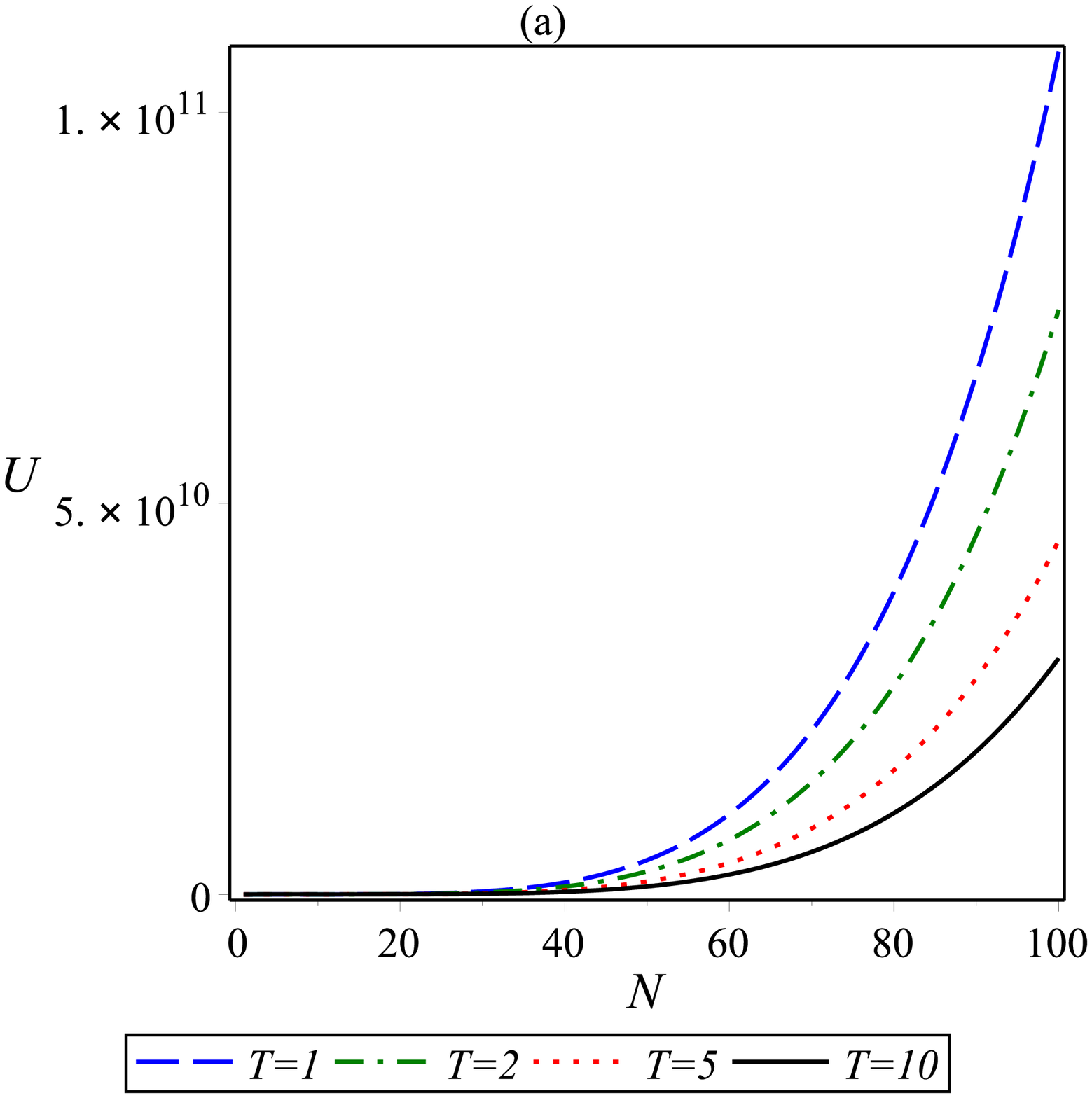}\includegraphics[width=65 mm]{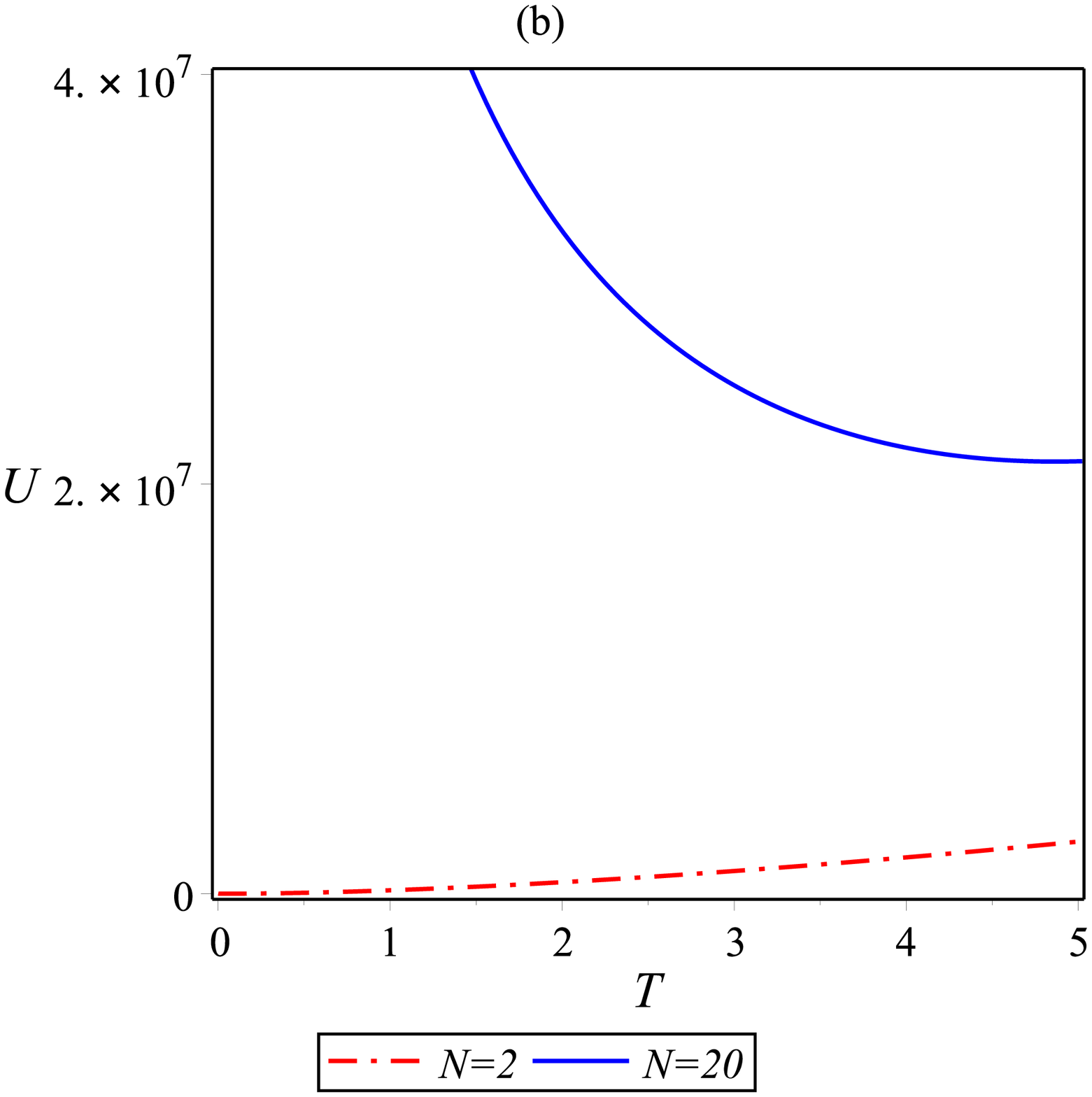}
 \end{array}$
 \end{center}
\caption{Energy of Tsallis statistics with $G=m=k=1$. (a) in terms of $N$; (b) in terms of $T$.}
 \label{fig9}
\end{figure}

\nd Finally, as is usual, the specific heat is obtained by:
\begin{equation}
\label{eq8.11}
C_{v}=\frac {\partial <U>} {\partial T}.
\end{equation}
Our numerical analysis indicated that specific heat diverges at special $N$, which is something like phase transition, hence we can suggest a maximum number for $N$ where the system is in thermodynamics stability. Also we can see that specific heat has a maximum in the stable region which is like Schottky anomaly. These are illustrated by Fig. \ref{fig7} (a). On the other hand, from Fig. \ref{fig7} (b) we can see behavior of specific heat in terms of the temperature. Similar to the energy, we can see different behavior for small and large $N$. For the small numbers of galaxies we see the specific heat is increasing function of temperature as expected.
\begin{figure}[h!]
 \begin{center}$
 \begin{array}{cccc}
\includegraphics[width=65 mm]{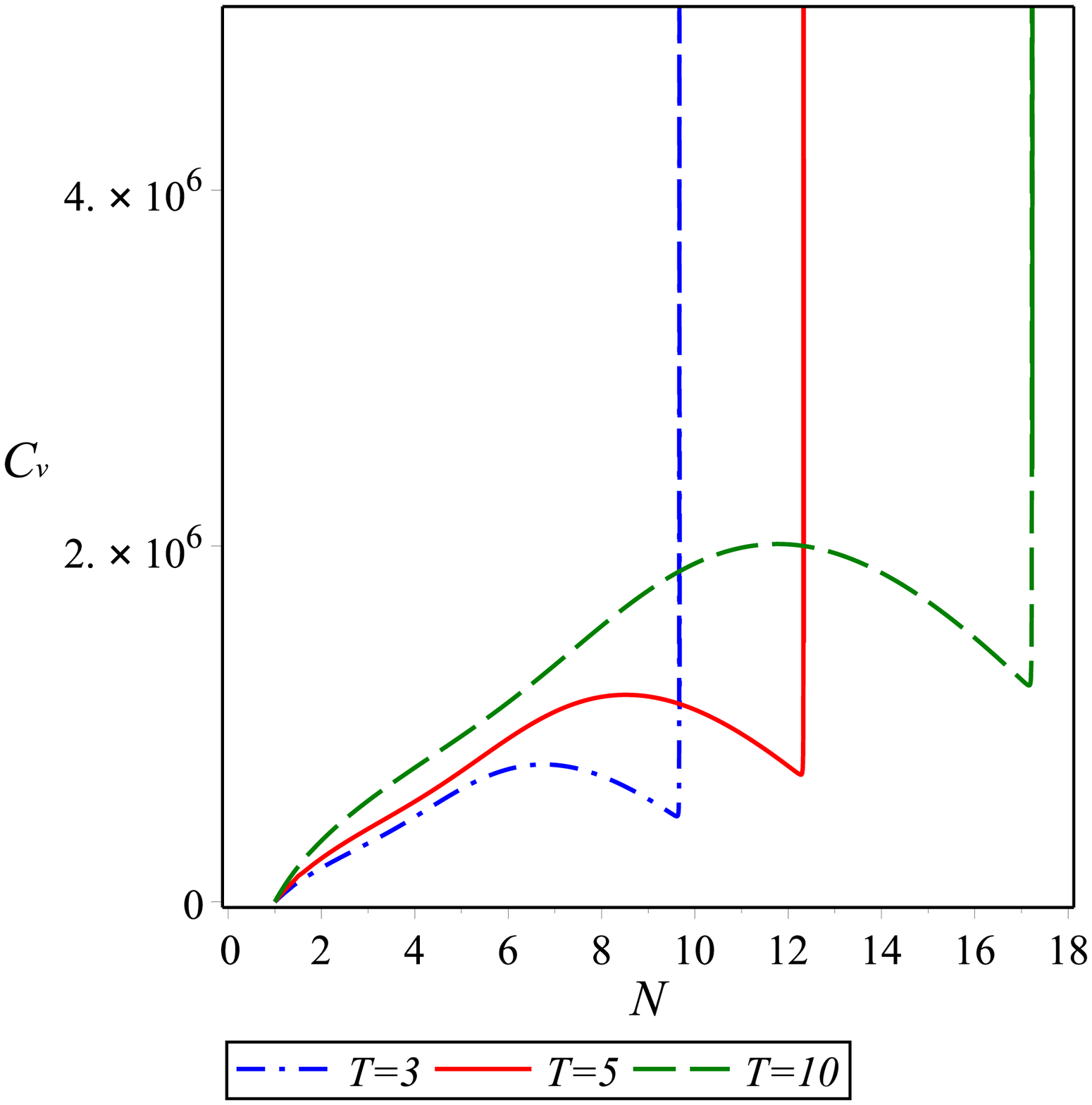}\includegraphics[width=65 mm]{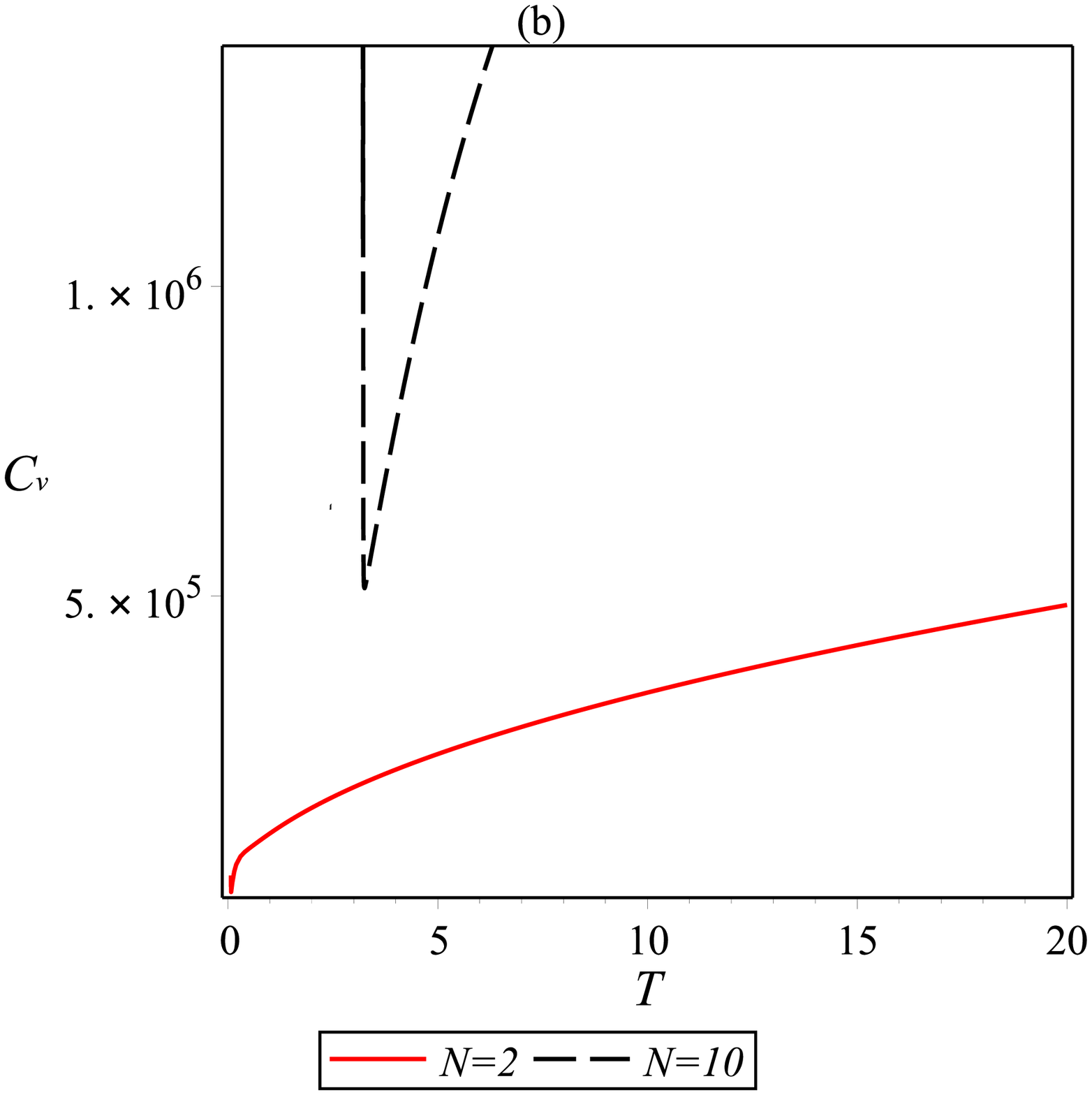}
 \end{array}$
 \end{center}
\caption{Specific heat of Tsallis statistics with $G=m=k=1$. (a) in terms of $N$; (b) in terms of $T$.}
 \label{fig10}
\end{figure}

\section{Conclusion}

The main objective of the paper is to obtain divergence free partition function for the system of galaxies clustering in the brane world model. Statistically adequate in content, the paper follows Boltzmann statistical and the Tsallis statistical treatment to evaluate the partition function and the corresponding thermodynamics in the form of equations of state. The speciality of paper lies in the successful elimination of divergences arising due to the point mass approximations, which otherwise is mathematically a challenging issue for cosmologists. In the present work the main conclusion of the paper is in the form of finite and singularity free partition function and thus the regularized thermodynamics. This task is accomplished by using the generalization of the dimensional regularization of Bollini and Giambiagi \cite{xz12, xz14, prd}. This generalization was based on the general quantification method of QFT's \cite{la12, la14, la16, la18} using Ultradistributions of Sebastiao
e Silva, also known as Ultrahyperfunctions \cite{jss, hasumi, tp8}. In 2018 the problem of exponentially divergent nature of gravitational partition function was solved by Plastino and Rocca \cite {di14, jpco, xz16}, thus deserves a special mention in the conclusive section of this paper.

\nd
General relativity has been thoroughly tested at solar system scale \cite{50}, and it is known to be valid at that scale. Thus, at that  scale the non-locality is  strongly constrained as general relativity is a local theory \cite{50a, 50b}. However, it is still possible for non-locality to occur at very large distance scales. In fact, it has been suggested that   quantum gravitational effects would modify the gravitational potential at large distances, and this can explain the accelerated expansion of the universe without dark energy  \cite{50c, 50d}. So, non-localities can have important consequences for large scale structure formation in our universe \cite{50ab}. Such non-local effects of gravity could be detected using  gravitational waves \cite{50ac}. It would be interesting to construct a non-local gravitational partition function, using the methods used in this paper.

\newpage

\renewcommand{\thesection}{\Alph{section}}

\renewcommand{\theequation}{\Alph{section}.\arabic{equation}}

\appendix

\section{Appendix: Auxiliary calculations}

\subsection{Calculations of: Tsallis partition function of the system in the brane world model}

\setcounter{equation}{0}

\nd
In 1988, C. Tsallis proposed a non-extensive generalization of statistical mechanics (widely called as Tsallis statistical mechanics) \cite{t1}. Based on the motivations laid out in the introduction, we now try to investigate the many-body gravitational system using Tsallis's formalism.
Tsallis q-exponential is defined as the distribution:
\begin{equation}
\label{eqa1}
e_q(x)=[1+(q-1)x]_+^{\frac {1} {q-1}}
\end{equation}
Or equivalently
\begin{equation}
\label{eqa.2}
e_q(x)=
\begin{cases}
1+(q-1)x]^{\frac {1} {q-1}}\;\;\;;\;\;\;1+(q-1)x>0\\
0\;\;\;;\;\;\;1+(q-1)x<0
\end{cases}
\end{equation}
Let's consider again the distribution:
$\frac {1} {r}=PV\frac {1} {r}$.
First we do the calculation in $\nu$ dimensions and assume  $q>1$. Thus
\begin{equation}
\label{eqa.3}
{\cal Z}_\nu=\int\limits_{-\infty}^{\infty}d^\nu x\int\limits_{-\infty}^{\infty}d^\nu p
\left[1+(q-1)\beta\left(\frac {N(N-1)Gm^2} {2r}+\frac{kN(N-1)Gm^2}{2r^3}-\frac {Np^2} {2m}\right)\right]_+^{\frac {1} {q-1}}
\end{equation}
Here we put $q=4/3$. The choice of that value of $q$ is not arbitrary. It is the value of $q$ where Verlinde's conjecture of emergent gravity could be proved, in the non-relativistic case \cite{di14}. Accordingly:
\begin{equation}
\label{eqa.4}
{\cal Z}_\nu=\left[\frac {2\pi^{\frac {\nu} {2}}} {\Gamma\left(\frac {\nu} {2}\right)}\right]^2
\int\limits_0^{\infty}r^{\nu-1}dr\int\limits_0^{\infty}p^{\nu-1} dp
\left[1+\beta\left(\frac {N(N-1)Gm^2} {6r}+\frac{N(N-1)Gm^2}{6r^3}-\frac {Np^2} {6m}\right)\right]_+^{3}
\end{equation}
The expression for ${\cal Z}$ in (\ref{eq5.4}) can be rewritten in the form:
\begin{equation}
\label{eqa.5}
{\cal Z}_\nu=\left[\frac {2\pi^{\frac {\nu} {2}}} {\Gamma\left(\frac {\nu} {2}\right)}\right]^2\left(\frac{6m}{\beta N}\right)^{\frac{\nu}{2}}B(4,\frac{\nu}{2})\int\limits_{0}^{\infty}r^{-\frac{\nu}{2}-10}dr\left(r^3+\alpha_t r+k\alpha_t\right)^{\frac{\nu}{2}+3}
\end{equation}
where
\[\alpha_t=\frac{\beta N(N-1)Gm^2}{6}\]
Hence we can write
\begin{equation}
\label{eqa.6}
{\cal Z}_\nu=\left[\frac {2\pi^{\frac {\nu} {2}}} {\Gamma\left(\frac {\nu} {2}\right)}\right]^2\left(\frac{6m}{\beta N}\right)^{\frac{\nu}{2}}B(4,\frac{\nu}{2})\int\limits_{0}^{\infty}r^{-\frac{\nu}{2}-10}dr(r+b)^{\frac{\nu}{2}+3}(r-r_1)^{\frac{\nu}{2}+3}(r-r_3)^{\frac{\nu}{2}+3}
\end{equation}
where $r_1$, $r_2=-b$, and $r_3$  are the cube roots of the integrand.
Also,
\[r_1=\gamma_1+e^{\frac{i\pi}{3}}\gamma_2\]
\[r_3=\gamma_1+e^{-\frac{i\pi}{3}}\gamma_2\]
\[b=\gamma_2-\gamma_1-\frac{\alpha}{3}\]
\[\gamma_1=\left(\sqrt{\frac{1}{4}\left(\frac{2\alpha_t^3}{27}+\alpha_t k\right)^2+\frac{1}{27}\left(\frac{\alpha}{3}(\alpha_t-6)\right)^2}-\frac{1}{2}\left(\frac{2\alpha_t^3}{27}+\alpha_t k\right)\right)^{\frac{1}{3}}\]
\[\gamma_2=\left(\sqrt{\frac{1}{4}\left(\frac{2\alpha_t^3}{27}+\alpha_t k\right)^2+\frac{1}{27}\left(\frac{\alpha}{3}(\alpha_t-6)\right)^2}+\frac{1}{2}\left(\frac{2\alpha_t^3}{27}+\alpha_t k\right)\right)^{\frac{1}{3}}\]
After solving the integral we have

\begin{eqnarray}\label{eq5a7}
{\cal Z}_\nu&=&\left[\frac {2\pi^{\frac {\nu} {2}}} {\Gamma\left(\frac {\nu} {2}\right)}\right]^2\left(\frac{6m}{\beta N}\right)^{\frac{\nu}{2}}B(4,\frac{\nu}{2})b^{\nu}\frac{\Gamma\left(-\frac{\nu}{2}-9\right)}{\Gamma\left(-\frac{3\nu}{2}-9\right)}\nonumber\\
&\times&\Gamma(-\nu){}F\left(-\nu,-\frac{\nu}{2}-3,-\frac{\nu}{2}-3,-\frac{3\nu}{2}-9;1+\frac{r_1}{b};1+\frac{r_3}{b}\right).
\end{eqnarray}
Here, $F$ is the hypergeometric function of two variables.
Using transformation formula we put $z_1=1+\frac{r_1}{b}$, $z_2=1+\frac{r_3}{b}$ and get:
\[\Gamma(-\nu){}F\left(-\nu,-\frac{\nu}{2}-3,-\frac{\nu}{2}-3,-\frac{3\nu}{2}-9;1+\frac{r_1}{b};1+\frac{r_3}{b}\right)=\]
\[\Gamma(-\nu)\biggl[1-\nu\left(\frac{-\frac{\nu}{2}-3}{-\frac{3\nu}{2}-9}z_1+\frac{-\frac{\nu}{2}-3}{-\frac{3\nu}{2}-9}z_2\right)\]
\[-\nu(1-\nu)\left(\frac{(-\frac{\nu}{2}-3)(-\frac{\nu}{2}-2)}{(-\frac{3\nu}{2}-9)(-\frac{3\nu}{2}-8)}z_1z_2+\frac{(-\frac{\nu}{2}-3)(-\frac{\nu}{2}-2)}{2(-\frac{3\nu}{2}-9)(-\frac{3\nu}{2}-8)}z_1^2+\frac{(-\frac{\nu}{2}-3)(-\frac{\nu}{2}-2)}{2(-\frac{3\nu}{2}-9)(-\frac{3\nu}{2}-8)}z_2^2\right)\]
\[-\nu(1-\nu)(2-\nu)\biggl(\frac{(-\frac{\nu}{2}-3)(-\frac{\nu}{2}-2)(-\frac{\nu}{2}-1)}{6(-\frac{3\nu}{2}-9)(-\frac{3\nu}{2}-8)(-\frac{3\nu}{2}-7)}z_1^3+\frac{(-\frac{\nu}{2}-3)(-\frac{\nu}{2}-2)(-\frac{\nu}{2}-1)}{6(-\frac{3\nu}{2}-9)(-\frac{3\nu}{2}-8)(-\frac{3\nu}{2}-7)}z_2^3+\]
\begin{equation}
\label{eqa.8}
\frac{(-\frac{\nu}{2}-3)(-\frac{\nu}{2}-2)(-\frac{\nu}{2}-3)}{2(-\frac{3\nu}{2}-9)(-\frac{3\nu}{2}-8)(-\frac{3\nu}{2}-7)}z_1^2z_1+\frac{(-\frac{\nu}{2}-3)(-\frac{\nu}{2}-2)(-\frac{\nu}{2}-3)}{2(-\frac{3\nu}{2}-9)(-\frac{3\nu}{2}-8)(-\frac{3\nu}{2}-7)}z_1z_2^2\biggr)\biggr]+\phi(z_1,z_2)
\end{equation}
where:
\begin{equation}
\label{eqa.9}
\phi_\nu(z_1,z_2)=\sum_{m+n=4}^\infty\frac{\Gamma\left(m+n-\nu\right)(-\frac{\nu}{2}-3)_m(-\frac{\nu}{2}-3)_n}{(-\frac{3\nu}{2}-9)_{m+n}m!n!}z_1^mz_2^n
\end{equation}
Equation 5.8 can be simplified and translated into the form:
\[\Gamma(-\nu){}F\left(-\nu,-\frac{\nu}{2}-3,-\frac{\nu}{2}-3,-\frac{3\nu}{2}-9;1+\frac{r_1}{b};1+\frac{r_3}{b}\right)=\]
\[\Gamma(-\nu)\biggl[1-\nu\left(\frac{-\frac{\nu}{2}-3}{-\frac{3\nu}{2}-9}\right)\left(z_1+z_2\right)
-\nu(1-\nu)\left(\frac{(-\frac{\nu}{2}-3)}{(-\frac{3\nu}{2}-9)(-\frac{3\nu}{2}-8)}\right)\]\[\left(2(-\frac{\nu}{2}-3)z_1z_2+(-\frac{\nu}{2}-2)z_1^2+(-\frac{\nu}{2}-2)z_2^2\right)\]
\[-\nu(1-\nu)(2-\nu)\biggl(\frac{(-\frac{\nu}{2}-3)(-\frac{\nu}{2}-2)}{6(-\frac{3\nu}{2}-9)(-3\frac{3\nu}{2}-8)(-\frac{3\nu}{2}-7)}\biggr)\]
\begin{equation}
\label{eqa.10}
\biggl((-\frac{\nu}{2}-1)z_1^3+(-\frac{\nu}{2}-1)z_2^3+3(-\frac{\nu}{2}-3)z_1^2z_1+3(-\frac{\nu}{2}-3)z_1z_2^2\biggr)\biggr]+\phi_\nu(z_1,z_2)
\end{equation}
Then, we have for $\nu=3$, the expression:
\begin{equation}
\label{eqa.11}
\phi_3(z_1,z_2)=\sum_{m+n=4}^\infty\frac{\Gamma\left(m+n-3\right)(-\frac{9}{2})_m(-\frac{9}{2})_n}{(-\frac{27}{2})_{m+n}m!n!}z_1^mz_2^n
\end{equation}

\[\Gamma(-\nu){}F\left(-\nu,-\frac{\nu}{2}-3,-\frac{\nu}{2}-3,-\frac{3\nu}{2}-9;1+\frac{r_1}{b};1+\frac{r_3}{b}\right)=\]
\[\Gamma(3-\nu)\biggl[\frac{1}{-\nu(1-\nu)(2-\nu)}+\left(\frac{-\frac{\nu}{2}-3}{(1-\nu)(2-\nu)(-\frac{3\nu}{2}-9)}\right)\left(z_1+z_2\right)
+\left(\frac{(-\frac{\nu}{2}-3)}{(2-\nu)(-\frac{3\nu}{2}-9)(-\frac{3\nu}{2}-8)}\right)\]\[\left(2(-\frac{\nu}{2}-3)z_1z_2+(-\frac{\nu}{2}-2)z_1^2+(-\frac{\nu}{2}-2)z_2^2\right)+\biggl(\frac{(-\frac{\nu}{2}-3)(-\frac{\nu}{2}-2)}{6(-\frac{3\nu}{2}-9)(-\frac{3\nu}{2}-8)(-\frac{3\nu}{2}-7)}\biggr)\]
\begin{equation}
\label{eqa.12}
\biggl((-\frac{\nu}{2}-1)z_1^3+(-\frac{\nu}{2}-1)z_2^3+3(-\frac{\nu}{2}-3)z_1^2z_1+3(-\frac{\nu}{2}-3)z_1z_2^2\biggr)\biggr]+\phi_\nu(z_1,z_2)
\end{equation}
Let us define $h_\nu$ such that

\[h_\nu=\biggl[\frac{1}{-\nu(1-\nu)(2-\nu)}+\left(\frac{-\frac{\nu}{2}-3}{(1-\nu)(2-\nu)(-\frac{3\nu}{2}-9)}\right)\left(z_1+z_2\right)
+\left(\frac{(-\frac{\nu}{2}-3)}{(2-\nu)(-\frac{3\nu}{2}-9)(-\frac{3\nu}{2}-8)}\right)\]\[\left(2(-\frac{\nu}{2}-3)z_1z_2+(-\frac{\nu}{2}-2)z_1^2+(-\frac{\nu}{2}-2)z_2^2\right)+\biggl(\frac{(-\frac{\nu}{2}-3)(-\frac{\nu}{2}-2)}{6(-\frac{3\nu}{2}-9)(-\frac{3\nu}{2}-8)(-\frac{3\nu}{2}-7)}\biggr)\]
\begin{equation}
\label{eqa.13}
\biggl((-\frac{\nu}{2}-1)z_1^3+(-\frac{\nu}{2}-1)z_2^3+3(-\frac{\nu}{2}-3)z_1^2z_1+3(-\frac{\nu}{2}-3)z_1z_2^2\biggr)\biggr]
\end{equation}
Thus we can write the partition function in $\nu$ dimensions as
\[{\cal Z}_\nu=\left[\frac {2\pi^{\frac {\nu} {2}}} {\Gamma\left(\frac {\nu} {2}\right)}\right]^2\left(\frac{6m}{\beta N}\right)^{\frac{\nu}{2}}B(4,\frac{\nu}{2})b^{\nu}\frac{\Gamma\left(-\frac{\nu}{2}-9\right)}{\Gamma\left(-\frac{3\nu}{2}-9\right)}\]
\begin{equation}
\label{eqa.14}
\left(\Gamma(3-\nu)h_\nu+\phi_\nu(z_1,z_2)\right)
\end{equation}
\[{\cal Z}_\nu=\frac {24\pi^\nu} {\Gamma\left(\frac {\nu} {2}\right)^2}\left(\frac{6m}{\beta N}\right)^{\frac{\nu}{2}}b^{\nu}\frac{\Gamma\left(-\frac{\nu}{2}-9\right)}{(3+\frac{\nu}{2})(2+\frac{\nu}{2})(1+\frac{\nu}{2})\Gamma\left(-\frac{3\nu}{2}-9\right)}\]
\begin{equation}
\label{eqa.15}
\left(\Gamma(3-\nu)h_\nu+\phi_\nu(z_1,z_2)\right)
\end{equation}

\subsection{Calculations of: Mean Energy in $\nu$ dimensions}

We now proceed to evaluate the average energy of the system. For it is:
\[{\cal Z}<{\cal U}>_\nu=\left[\frac {2\pi^{\frac {\nu} {2}}} {\Gamma\left(\frac {\nu} {2}\right)}\right]^2
\int\limits_0^{\infty}r^{\nu-1}dr
\int\limits_0^{P_0} p^{\nu-1} dp
\left(-\frac {N(N-1)Gm^2} {2r}-\frac {kN(N-1)Gm^2} {2r^2}+\frac {Np^2} {2m}\right)\]
\begin{equation}
\label{eqa.16}
\left[1+(q-1)\beta\left(\frac {N(N-1)Gm^2} {2r}+\frac{kN(N-1)Gm^2}{2r^3}-\frac{Np^2}{2m}\right)\right]^{\frac {1} {q-1}}
\end{equation}
Or, equivalently:
\[{\cal Z}<U>_\nu=\biggl[\frac{2\pi^\frac{\nu}{2}}{\Gamma{\frac{\nu}{2}}}\biggr]^2[\frac{\beta N(q-1)}{2m}]^{\frac{1}{q-1}}\]
\[\left(\int\limits_0^{\infty}r^{\nu-1}dr
\int\limits_0^{P_0}\frac{Np^{\nu+1}}{2m}\left[\frac{1}{\beta N(q-1)}+\frac{(N-1)Gm^2}{2r}+\frac{k(N-1)Gm^2}{2r^3}-\frac{p^2}{2m}\right]^{\frac{1}{q-1}}dp\right.\]
\[\left.-\frac {N(N-1)Gm^2} {2}\int\limits_0^{\infty}r^{\nu-2}dr\int\limits_0^{P_0}p^{\nu-1}\left[\frac{1}{\beta N(q-1)}+\frac{(N-1)Gm^2}{2r}+\frac{k(N-1)Gm^2}{2r^3}-\frac{p^2}{2m}\right]^{\frac{1}{q-1}}dp\right)\]
\begin{equation}
\label{eqa.17}
\left.-\frac {kN(N-1)Gm^2} {2}\int\limits_0^{\infty}r^{\nu-3}dr\int\limits_0^{P_0}p^{\nu-1}\left[\frac{1}{\beta N(q-1)}+\frac{(N-1)Gm^2}{2r}+\frac{k(N-1)Gm^2}{2r^3}-\frac{p^2}{2m}\right]^{\frac{1}{q-1}}dp\right)
\end{equation}
For $q=4/3$ we obtain:
\begin{equation}
\label{eqa.18}
{\cal Z}<U>_\nu=\frac {3} {2\beta}\left[\frac {2\pi^{\frac {\nu} {2}}} {\Gamma\left(\frac {\nu} {2}\right)}\right]^2
\left(\frac {6m} {N\beta}\right)^{\frac {\nu} {2}}
\biggl[-\alpha B(4,\frac{\nu}{2})I_1-k\alpha B(4,\frac{\nu}{2})I_2+B(4,\frac{\nu}{2}+1)I_3\biggr]
\end{equation}
where
\begin{equation}
\label{eqa.19}
I_1=\int\limits_0^{\infty}r^{-\frac{\nu}{2}-11}dr(r^3+\alpha r^2+k\alpha)^{3+\frac{\nu}{2}}
\end{equation}
\begin{equation}
\label{eqa.20}
I_2=\int\limits_0^{\infty}r^{-\frac{\nu}{2}-13}dr(r^3+\alpha r^2+k\alpha)^{3+\frac{\nu}{2}}
\end{equation}
\begin{equation}
\label{eqa.21}
I_3=\int\limits_0^{\infty}r^{-\frac{\nu}{2}-13}dr(r^3+\alpha r^2+k\alpha)^{4+\frac{\nu}{2}}
\end{equation}
we can further write the integrals as
\begin{equation}
\label{eqa.22}
I_1=\int\limits_{0}^{\infty}r^{-\frac{\nu}{2}-11}dr(r+b)^{\frac{\nu}{2}+3}(r-r_1)^{\frac{\nu}{2}+3}(r-r_3)^{\frac{\nu}{2}+3}
\end{equation}
\begin{equation}
\label{eqa.23}
I_2=\int\limits_{0}^{\infty}r^{-\frac{\nu}{2}-13}dr(r+b)^{\frac{\nu}{2}+3}(r-r_1)^{\frac{\nu}{2}+3}(r-r_3)^{\frac{\nu}{2}+3}
\end{equation}

\begin{equation}
\label{eqa.24}
I_3=\int\limits_{0}^{\infty}r^{-\frac{\nu}{2}-13}dr(r+b)^{\frac{\nu}{2}+4}(r-r_1)^{\frac{\nu}{2}+4}(r-r_3)^{\frac{\nu}{2}+4}
\end{equation}
Thus after using the formula given below we evaluate the integrals

\[\int\limits_{0}^{\infty}r^{\gamma-\alpha-1}dr(r+b)^{B+B^{'}-\gamma}(r-r_1)^{B}(r-r_3)^{B^{'}}=b^{-\alpha}\frac{\Gamma\left(\gamma-\alpha\right)}{\Gamma\left(\gamma\right)}\]
\begin{equation}
\label{eqa.25}
\Gamma(\alpha){}F\left(\alpha,B,B^{'},\gamma;1+\frac{r_1}{b};1+\frac{r_3}{b}\right)
\end{equation}
Thus we get:
\[I_1=b^{\nu}\frac{\Gamma\left(-\frac{\nu}{2}-10\right)}{\Gamma\left(-\frac{3\nu}{2}-10\right)}\]
\begin{equation}
\label{eqa.26}
\Gamma(-\nu){}F\left(-\nu,-\frac{\nu}{2}-3,-\frac{\nu}{2}-3,-\frac{3\nu}{2}-10;1+\frac{r_1}{b};1+\frac{r_3}{b}\right)
\end{equation}
\[I_2=b^{\nu}\frac{\Gamma\left(-\frac{\nu}{2}-12\right)}{\Gamma\left(-\frac{3\nu}{2}-12\right)}\]
\begin{equation}
\label{eqa.27}
\Gamma(-\nu){}F\left(-\nu,-\frac{\nu}{2}-3,-\frac{\nu}{2}-3,-\frac{3\nu}{2}-12;1+\frac{r_1}{b};1+\frac{r_3}{b}\right)
\end{equation}
\[I_3=b^{\nu}\frac{\Gamma\left(-\frac{\nu}{2}-12\right)}{\Gamma\left(-\frac{3\nu}{2}-12\right)}\]
\begin{equation}
\label{eqa.28}
\Gamma(-\nu){}F\left(-\nu,-\frac{\nu}{2}-4,-\frac{\nu}{2}-4,-\frac{3\nu}{2}-12;1+\frac{r_1}{b};1+\frac{r_3}{b}\right)
\end{equation}
Using transformation formula
\begin{equation}
\label{eqa.29}
\Gamma(-\nu){}F\left(-\nu,-\frac{\nu}{2}-3,-\frac{\nu}{2}-3,-\frac{3\nu}{2}-10;1+\frac{r_1}{b};1+\frac{r_3}{b}\right)=\Gamma(3-\nu)h_1+\phi_1(\nu,z_1,z_2)
\end{equation}
Where we put
\[h_1(\nu)=\biggl[\frac{1}{-\nu(1-\nu)(2-\nu)}+\left(\frac{-\frac{\nu}{2}-3}{(1-\nu)(2-\nu)(-\frac{3\nu}{2}-10)}\right)\left(z_1+z_2\right)+\]
\[\left(\frac{(-\frac{\nu}{2}-3)}{(2-\nu)(-\frac{3\nu}{2}-10)(-\frac{3\nu}{2}-9)}\right)\times\]\[\left(2(-\frac{\nu}{2}-3)z_1z_2+(-\frac{\nu}{2}-2)z_1^2+(-\frac{\nu}{2}-2)z_2^2\right)+\biggl(\frac{(-\frac{\nu}{2}-3)(-\frac{\nu}{2}-2)}{6(-\frac{3\nu}{2}-10)(-\frac{3\nu}{2}-9)(-\frac{3\nu}{2}-8)}\biggr)\]
\begin{equation}
\label{eqa.30}
\biggl((-\frac{\nu}{2}-1)z_1^3+(-\frac{\nu}{2}-1)z_2^3+3(-\frac{\nu}{2}-3)z_1^2z_2+3(-\frac{\nu}{2}-3)z_1z_2^2\biggr)\biggr]
\end{equation}
In three dimensions:
\begin{equation}
\label{eqa.31}
h_1(3)=\biggl[-\frac{1}{6}+\frac{9}{58}(z_1+z_2)-\frac{18}{87}z_1z_2+\frac{7}{87}(z_1^2+z_2^2)+\frac{7}{2610}(z_1^3+z_2^3)+\frac{7}{50}(z_1^2z_2+z_1z_2^2)\biggr]
\end{equation}
And:
\begin{equation}
\label{eqa.32}
[h_1^{'}(3)=\biggl[\frac{11}{36}-\frac{779}{3364}(z_1+z_2)+\frac{518}{2523}z_1z_2+\frac{65}{841}(z_1^2+z_2^2)+\frac{101}{315375}(z_1^3+z_2^3)+\frac{697}{1576875}(z_1^2z_2+z_1z_2^2)\biggr]
\end{equation}
In the same way:
\begin{equation}
\label{eqa.33}
\Gamma(-\nu){}F\left(-\nu,-\frac{\nu}{2}-3,-\frac{\nu}{2}-3,-\frac{3\nu}{2}-12;1+\frac{r_1}{b};1+\frac{r_3}{b}\right)=\Gamma(3-\nu)h_2+\phi_2(\nu,z_1,z_2)
\end{equation}
Where:
\[h_2(\nu)=\biggl[\frac{1}{-\nu(1-\nu)(2-\nu)}+\left(\frac{-\frac{\nu}{2}-3}{(1-\nu)(2-\nu)(-\frac{3\nu}{2}-12)}\right)\left(z_1+z_2\right)+\]
\[\left(\frac{(-\frac{\nu}{2}-3)}{(2-\nu)(-\frac{3\nu}{2}-12)(-\frac{3\nu}{2}-11)}\right)\times\]
\[\left(2(-\frac{\nu}{2}-3)z_1z_2+(-\frac{\nu}{2}-2)z_1^2+(-\frac{\nu}{2}-2)z_2^2\right)+\biggl(\frac{(-\frac{\nu}{2}-3)(-\frac{\nu}{2}-2)}{6(-\frac{3\nu}{2}-12)(-\frac{3\nu}{2}-11)(-\frac{3\nu}{2}-10)}\biggr)\]
\begin{equation}
\label{eqa.34}
\biggl((-\frac{\nu}{2}-1)z_1^3+(-\frac{\nu}{2}-1)z_2^3+
3(-\frac{\nu}{2}-3)z_1^2z_1+3(-\frac{\nu}{2}-3)z_1z_2^2\biggr)\biggr]
\end{equation}
Again, in three dimensions:
\begin{equation}
\label{eqa.35}
h_2(3)=\biggl[-\frac{1}{6}+\frac{3}{22}(z_1+z_2)-\frac{54}{341}z_1z_2+\frac{21}{341}(z_1^2+z_2^2)+\frac{35}{1798}(z_1^3+z_2^3)+\frac{189}{1798}(z_1^2z_2+z_1z_2^2)\biggr]
\end{equation}
And:
\[h_2^{'}(3)=\biggl[\frac{11}{36}+\frac{293}{1452}(z_1+z_2)-\frac{17778}{116281}z_1z_2+\frac{20059}{348843}(z_1^2+z_2^2)+\frac{461521}{880130889}(z_1^3+z_2^3)+\]
\begin{equation}
\label{eqa36}
\frac{69105}{97792321}(z_1^2z_2+z_1z_2^2)\biggr]
\end{equation}
The last equality is:
\begin{equation}
\label{eqa.37}
\Gamma(-\nu){}F\left(-\nu,-\frac{\nu}{2}-4,-\frac{\nu}{2}-4,-\frac{3\nu}{2}-12;1+\frac{r_1}{b};1+\frac{r_3}{b}\right)=\Gamma(3-\nu)h_3+\phi_3(\nu,z_1,z_2)
\end{equation}
With:
\[h_3(\nu)=\biggl[\frac{1}{-\nu(1-\nu)(2-\nu)}+\left(\frac{-\frac{\nu}{2}-4}{(1-\nu)(2-\nu)(-\frac{3\nu}{2}-12)}\right)\left(z_1+z_2\right)+\]
\[\left(\frac{(-\frac{\nu}{2}-4)}{(2-\nu)(-\frac{3\nu}{2}-12)(-\frac{3\nu}{2}-11)}\times\right)\]\[\left(2(-\frac{\nu}{2}-4)z_1z_2+(-\frac{\nu}{2}-3)z_1^2+(-\frac{\nu}{2}-3)z_2^2\right)+\biggl(\frac{(-\frac{\nu}{2}-4)(-\frac{\nu}{2}-3)}{6(-\frac{3\nu}{2}-12)(-\frac{3\nu}{2}-11)(-\frac{3\nu}{2}-10)}\biggr)\]
\begin{equation}
\label{eqa.38}
\biggl((-\frac{\nu}{2}-2)z_1^3+(-\frac{\nu}{2}-2)z_2^3+3(-\frac{\nu}{2}-4)z_1^2z_1+3(-\frac{\nu}{2}-4)z_1z_2^2\biggr)\biggr]
\end{equation}
In three dimensions:
\begin{equation}
\label{eqa.39}
h_3(3)=\biggl[-\frac{1}{6}+\frac{1}{6}(z_1+z_2)-\frac{22}{93}z_1z_2-\frac{3}{31}(z_1^2+z_2^2)+\frac{7}{1798}(z_1^3+z_2^3)+\frac{33}{1798}(z_1^2z_2+z_1z_2^2)\biggr]
\end{equation}
\begin{equation}
\label{eqa.40}
h_3^{'}(3)=\biggl[\frac{11}{36}-\frac{1}{4}(z_1+z_2)+\frac{686}{2883}z_1z_2+\frac{275}{2883}(z_1^2+z_2^2)+\frac{1522}{7273809}(z_1^3+z_2^3)+\frac{80}{2406623}(z_1^2z_2+z_1z_2^2)\biggr]
\end{equation}
We then obtain:
\begin{equation}
\label{eqa.41}
{\cal Z}<U>_\nu=G_\nu[\Gamma(3-\nu)F_\nu+\Phi_\nu(z_1,z_2)\biggr]
\end{equation}
Where we have defined:
\[\Phi_\nu(z_1,z_2)=\phi_1+\phi_2+\phi_3\]
With:
\begin{equation}
\label{eqa.42}
G_\nu=\frac{3}{2\beta}\frac {24\pi^\nu} {\Gamma\left(\frac {\nu} {2}\right)^2}\left(\frac{6m}{\beta N}\right)^{\frac{\nu}{2}}b^{\nu}\frac{\Gamma\left(\frac{3\nu}{2}+11\right)\sin{\frac{3\nu\pi}{2}}}{(3+\frac{\nu}{2})(2+\frac{\nu}{2})(1+\frac{\nu}{2})\Gamma\left(\frac{\nu}{2}+11\right)\sin{\frac{\pi\nu}{2}}}
\end{equation}
And:
\begin{equation}
\label{eqa.43}
F_\nu=\left[-\alpha h_1(\nu)-k\alpha h_2(\nu)\frac{(\frac{3\nu}{2}+12)(\frac{3\nu}{2}+11)}{(\frac{\nu}{2}+12)(\frac{\nu}{2}+11)}+h_3(\nu)\frac{(\frac{3\nu}{2}+12)(\frac{3\nu}{2}+11)}{(\frac{\nu}{2}+12)(\frac{\nu}{2}+11)}\right]
\end{equation}

\end{document}